\RequirePackage[2020-02-02]{latexrelease}%ESTA LINEA LA AGREGUE YA QUE NO COMPILABA EL DOCUMENTO EN MI VERSIÓN DE TeXStudio, si presenta problemas para compilar por favor solo marque la linea como comentario y no deberia existir ningun otro problema, NO BORRAR POR FAVOR.
\documentclass[showpacs,aps,prd,nofootinbib,floatfix,amsmath,amssymb,twocolumn]{revtex4}
\usepackage{mathrsfs}
\usepackage{graphicx}
\usepackage[dvipsnames]{xcolor}
\usepackage{dsfont}
\begin{document}

\makeatletter
%Feynman slash
\newbox\slashbox \setbox\slashbox=\hbox{$/$}
\newbox\Slashbox \setbox\Slashbox=\hbox{\large$/$}
\def\pFMslash#1{\setbox\@tempboxa=\hbox{$#1$}
  \@tempdima=0.5\wd\slashbox \advance\@tempdima 0.5\wd\@tempboxa
  \copy\slashbox \kern-\@tempdima \box\@tempboxa}
\def\pFMSlash#1{\setbox\@tempboxa=\hbox{$#1$}
  \@tempdima=0.5\wd\Slashbox \advance\@tempdima 0.5\wd\@tempboxa
  \copy\Slashbox \kern-\@tempdima \box\@tempboxa}
\def\FMslash{\protect\pFMslash}
\def\FMSlash{\protect\pFMSlash}
\def\miss#1{\ifmmode{/\mkern-11mu #1}\else{${/\mkern-11mu #1}$}\fi}
%%%% Uso:  \pFMSlash{p}
\makeatother

%\tightenlines
\title{Lepton-flavor changing decays and non-unitarity in the inverse seesaw mechanism}

\author{Adri\'an Gonz\'alez-Quiterio}
\author{H\'ector Novales-S\'anchez}
\affiliation{
Facultad de Ciencias F\'isico Matem\'aticas, Benem\'erita Universidad Aut\'onoma de Puebla, Apartado Postal 1152 Puebla, Puebla, M\'exico}

\begin{abstract}
	The pursuit for the genuine fundamental description, governing nature at some high-energy scale, must invariably consider the yet-unknown mechanism behind the generation of neutrino mass. Lepton-flavor violating decays $l_\alpha\to l_\beta\,\gamma$, allowed in the presence of neutrino mass and mixing, provide a mean to look for physics beyond the Standard Model. In the present work we consider the inverse seesaw mechanism and then revisit the calculation of its contributions to the branching ratios of the aforementioned decay processes. Our analytic results are consistent, as they are gauge invariant, gauge independent, ultraviolet finite, and of decoupling nature. Among the decays $l_\alpha\to l_\beta\gamma$, we find $\mu\to\gamma\,e$ to be the most promising, in the light of current bounds by the MEG Collaboration. Deviations from unitarity in the mixing of light neutrinos are related to the branching ratios ${\rm Br}\big( l_\alpha\to\gamma\,l_\beta \big)$ in a simple manner, which we address, then finding that the constraints on non-unitarity effects from MEG data will be improved by the upcoming MEG II update by a factor $\sim\frac{1}{3}$.
\end{abstract}

\pacs{14.60.Pq, 14.60.St, 13.35.$-$r}

\maketitle

\section{Introduction}
The key role played by gauge symmetry in the description of fundamental physics and the origin of mass through the occurrence of spontaneous symmetry breaking has been greatly supported by the measurement, by the ATLAS and CMS Collaborations at the CERN, of what seems to be the long-awaited Higgs boson~\cite{HiggsATLAS,HiggsCMS}. In the minimal theoretical scheme, provided by the scalar sector of the Standard Model~\cite{SMGlashow,SMSalam,SMWeinberg}, the Brout-Englert-Higgs mechanism breaks the electroweak gauge symmetry group ${\rm SU}(2)_L\otimes{\rm U}(1)_Y$ into the electromagnetic group ${\rm U}(1)_e$~\cite{EnBr,Higgs}, which comes along with the definition of the masses of all the currently measured particles, except for the neutrinos, assumed to be massless in this framework. Neutrino masses have not ever been measured~\cite{PDG}, and yet the widely accepted interpretation that neutrino oscillations~\cite{Pontecorvo} imply massiveness of these particles calls for an explanation emerging from some high-energy formulation of physics beyond the Standard Model. Though the addition of three sterile right-handed neutrino fields to the Standard-Model particle spectra, together with a set of Dirac-type Yukawa terms, does the job, the resulting neutrino masses are determined by {\it ad hoc} ``unnaturally'' small values of Yukawa constants. In view of the massiveness and electromagnetic neutrality of neutrinos, the proper description of these particles could rather correspond to Majorana spinor fields~\cite{Majorana}, in which case lepton number is not conserved. If lepton-number non-preservation is assumed, the effective Lagrangian for the electroweak Standard Model~\cite{LLR,BuWy,Wudka} gets extended, then allowing for the emergence of Majorana mass-terms~\cite{Weinbergoperator,BaLe,CSV}, presumably originating in some description of fundamental physics characterized by a high-energy scale, $\Lambda$. In particular, the Weinberg operator~\cite{Weinbergoperator} yields neutrino masses matching the mass profile that characterizes the seesaw mechanism~\cite{MoSe1,MoSe2}, in which, besides the known neutrinos, a set of heavy-neutral-lepton partners, dubbed ``heavy neutrinos'', comes about. The heavy-neutrino masses, $m_{N_j}$, determined by the high-energy scale as $m_{N_j}\sim\Lambda$, are linked to the masses $m_{\nu_j}$, of the light neutrinos, in this scheme given by $m_{\nu_j}\sim\frac{v^2}{\Lambda}$, with $v$ the vacuum expectation value of the Standard-Model Higgs potential. In this context, current upper bounds on light-neutrino masses~\cite{Planck,cosmonumass,KATRIN} push the masses of heavy neutrinos towards enormous values, thus leaving direct production of heavy neutrinos off the table and also severely attenuating their contributions, as virtual particles, to Standard-Model observables. 
\\

So even though the seesaw mechanism provides a nice explanation for the tininess of neutrino mass, in connection with fundamental physics beyond the Standard Model, it is quite difficult to probe. This inconvenience has motivated the realization of seesaw variants, aimed at a relaxation of the restriction on the energy scale $\Lambda$, in order to bring it closer to current experimental sensitivity. Ref.~\cite{CHLR} provides a review on the seesaw mechanism and its variants. The framework for the present paper is defined by the neutrino-mass generation approach known as the ``inverse seesaw mechanism''~\cite{MoVa,GoVa,DeVa}. More concretely, we consider a realization of the inverse seesaw in which the neutrino sector is enriched, as compared to the case of the Standard Model, by the introduction of three right-handed neutrino fields, together with a set of three further left-handed lepton fields, all of them singlets with respect to the ${\rm SU}(2)_L\otimes{\rm U}(1)_Y$ gauge-symmetry group. 
%Besides these fermion fields, a complex scalar field is also added, which causes an extension of the scalar sector. 
These new fermion fields introduce a slight violation of lepton number through Majorana-like mass terms characterized by two small-valued matrices, here denoted by $\mu_S$ and $\mu_R$, both proportional to an energy scale $v_\sigma$, of spontaneous symmetry breaking, which abides by the hierarchy condition $v_\sigma\ll v$. In the mass-eigenspinor basis, this extended neutrino sector yields three light neutrinos and a total of six heavy neutral leptons. The mass matrix of light neutrinos, $m_\nu$, is proportional to $\mu_S$, which therefore attenuates the pressure on $\Lambda$, thus allowing for smaller and more reasonable values of the heavy-neutrino masses, as compared to what happens in the original version of the seesaw mechanism. 
%Note that the smallness of $\mu_S$ and $\mu_R$, after quantum corrections, relies on naturalness arguments, in the 't Hooft sense~\cite{}.
\\

%In the mass-eigenspinor basis, this extended neutrino sector yields three light neutrinos and a total of six heavy neutral leptons. This course of events introduces two matrices, here denoted by $\mu_S$ and $\mu_R$, both proportional to an energy scale $v_\sigma$, of spontaneous symmetry breaking, which abide by the hierarchy condition $v_\sigma\ll v\ll\Lambda$, with $v$ the Standard-Model vacuum expectation value. After symmetry breaking at $v_\sigma$, a Goldstone boson, known as the ``majoron'', emerges from the new scalar field. Due to its origin, the majoron is massless, though works are available, such as Refs.~\cite{}, in which the majoron is endowed with mass, then rendering  this particle a viable dark matter candidate~\cite{}. The mass matrix of light neutrinos, $m_\nu$, is proportional to $\mu_S$, which therefore attenuates the pressure on $\Lambda$, thus allowing for smaller and more reasonable values of the heavy neutrinos, as compared to what happens in the original version of the seesaw mechanism. Note that the smallness of $\mu_S$ and $\mu_R$, after quantum corrections, relies on naturalness arguments, in the 't Hooft sense~\cite{}.\\

In the electroweak Standard Model, the absence of right-handed neutrinos and the assumption of lepton-number conservation prevent neutrino-mass terms from being generated. Moreover, lepton-flavor-violating processes, also forbidden in such a context, are well-motivated means to search for traces pointing towards the presence of new physics. Despite the large amount of experimental work dedicated to the search for charged-lepton-flavor violation, no process of this kind has ever been observed~\cite{PDG}, though notice that neutrino mixing, required for the occurrence of neutrino oscillations~\cite{Pontecorvo}, implies that such sort of processes are actually allowed, in which case their measurement could be eventually achieved in experimental facilities. The present investigation readdresses the one-loop contributions, generated by the whole set of massive neutral leptons defined in the context of the inverse seesaw, to the charged-lepton-flavor-changing decays $\mu\to e\,\gamma$, $\tau\to e\,\gamma$, and $\tau\to \mu\,\gamma$. These decay processes have been formerly examined in Refs.~\cite{SFLYC,SoRu,GPH}. In contrast with such previous works, our calculations and estimations of the branching ratios do not involve any approximation regarding the masses of light neutrinos nor those of charged leptons. Our work also includes a discussion on the behavior of the contributions as $\Lambda$ is assumed to be very large, from which we arrive at the conclusion that the new physics is of decoupling nature. Moreover, while in these previous studies the scale of lepton number violation is assumed to lie within the KeV regime, for our estimations we have considered a larger set of energies for this scale, which we run from the few KeVs up to energies of a few MeVs. Also, both degenerate and non-degenerate sets of masses for the heavy-neutral leptons have been used to estimate the contributions. Currently, the most stringent experimental constraints on these decay processes are the ones reported by the MEG, Belle, and BaBar collaborations~\cite{MEGlfvbound,Bellelfvbound,BaBarlfvbound}. The MEG~II and Belle~II upgrades, expected to improve experimental sensitivities by $\sim1$ order of magnitude~\cite{MEG2lfvestimation,Belle2lfvestimation}, are also to be borne in mind. Our estimations show that, in accordance with such experimental works, the inverse-seesaw contribution to $\mu\to e\,\gamma$ yields the most promising result among the charged-lepton-flavor-violating decay processes under consideration. A worthwhile aspect of the inverse seesaw regards the matrix that characterizes the mixing of light neutrinos with charged leptons. In the simplest scenarios, as the Standard Model endowed with three singlet right-handed Dirac neutrino fields~\cite{GiKi} or the one given by the introduction of the Weinberg operator with the assumption that lepton number is not preserved~\cite{Weinbergoperator}, light-neutrino mixing is characterized by the Pontecorvo-Maki-Nakgawa-Sakata (PMNS) matrix~\cite{MNSmatrix,Pontecorvomatrix}, which is unitary. However, the presence of a light-neutrino mixing matrix which is not unitary occurs in a large class of models aimed at generating neutrino masses~\cite{FGLY}. Constraints imposed by the charged-lepton-flavor-violating decays $\mu\to e\,\gamma$, $\tau\to e\,\gamma$, and $\tau\to \mu\,\gamma$ on these non-unitary effects have been estimated in Refs.~\cite{FHL,BFHLMN}. In the present work, we analyze the aforementioned lepton-flavor-changing decays in terms of their relation with such non-unitary effects. The most restrictive constraints proceed from the contributions to $\mu\to\,e\gamma$ and its comparison with the MEG limit. Furthermore, we find that the sensitivity expectation estimated for MEG II should yield a potential improvement, given by a factor $\sim\frac{1}{3}$, of restrictions for the non-unitary effects.
\\

The reminder of the paper has been organized as follows: in Section~\ref{theory}, the inverse seesaw mechanism is reviewed, which includes the main expressions for the phenomenological calculation to be carried out later; we describe the execution of our analytical calculation of the charged-lepton-flavor-violating decays  $\mu\to e\,\gamma$, $\tau\to e\,\gamma$, and $\tau\to \mu\,\gamma$ at one loop in Section~\ref{analytic}; our estimations and analyses of the contributions are developed throughout Section~\ref{numbers}; we conclude the paper by providing, in Section~\ref{summary}, a summary. 
\\

%%%%%%%%%%%%%%%%%%%%
%%%%%%%%%%%%%%%%%%%%
%%%%%%%%%%%%%%%%%%%%
%%%%%%%%%%%%%%%%%%%%
%%%%%%%%%%%%%%%%%%%%

\section{The inverse seesaw mechanism}
\label{theory}
In this section, we discuss the inverse seesaw mechanism~\cite{MoVa,GoVa,DeVa}, which is the framework behind our phenomenological investigation. 
%Aiming at a more definite presentation, we address this task by considering a completion in which lepton number is spontaneously broken~\cite{CMP}. 
Think of an extension of the Standard Model, characterized by a Lagrangian invariant under the electroweak gauge group ${\rm SU}(2)_L\otimes{\rm U}(1)_Y$. Note that such a description might originate from some fundamental theory, after the occurrence of a stage of spontaneous symmetry breaking at some high-energy scale $\Lambda$. We work under the premise that $v\ll\Lambda$, where $v=246\,{\rm GeV}$ is the vacuum expectation value of the Higgs potential. Assume that the Standard-Model field content has been increased by the introduction of a set of three chiral fermion fields $\nu_{1,R}$, $\nu_{2,R}$, and $\nu_{3,R}$, which are right handed with lepton number $L(\nu_R)=+1$, and by three more fermion fields $S_{1,L}$, $S_{2,L}$, and $S_{3,L}$, with definite left chirality and lepton number $L(S_L)=+1$. All these fields are assumed to be singlets with respect to electroweak gauge transformations. 
%Then, a complex scalar field, $\sigma$, assumed to be an ${\rm SU}(2)_L\otimes{\rm U}(1)_Y$ singlet and to have lepton  number $L(\sigma)=+2$, is introduced as well. 
The addition of these new-physics fields translates into an extension of the lepton-Yukawa sector of the Standard Model, here denoted by ${\cal L}_{\rm Y}$. 
We assume that the expression of the Yukawa-sector Lagrangian, after the breaking of the ${\rm SU}(2)_L\otimes{\rm U}(1)_Y$ gauge symmetry, at $v$,
%and then using $\sigma=\frac{v_\sigma}{\sqrt{2}}+\cdots$, to break the global ${\rm U}(1)$ group, 
is
\begin{eqnarray}
&&
{\cal L}_{\rm Y}=\sum_{\alpha=e,\mu,\tau}\sum_{k=1}^3\Big(-(m_{\rm D})_{\alpha k}\,\overline{\nu_{\alpha,L}}\,\nu_{k,R}+{\rm H.c.} \Big)
\nonumber \\ &&\hspace{0.8cm}
+\sum_{j=1}^3\sum_{k=1}^3\Big(-M_{jk}\overline{S_{j,L}}\nu_{k,R}
%\nonumber \\ &&\hspace{0.3cm}
-\frac{(\mu_S)_{jk}}{2}\overline{S_{j,L}}S_{k,L}^{\rm c}
\nonumber \\ &&\hspace{0.8cm}
-\frac{(\mu_R)_{jk}}{2}\overline{\nu_{j,R}^{\rm c}}\nu_{k,R}+{\rm H.c.} \Big)+\cdots.
\nonumber \\
\label{Yukawamass}
\end{eqnarray}
where the ellipsis denote further Yukawa terms, including those made exclusively of Standard-Model fields.
%By considering the left-handed unitary-transformation matrix $V^\ell_L$, we found it convenient to define $\nu_L=V^{\ell\dag}_L\nu'_L$. %Then the first term in the right-hand side of this equation features the $3\times3$ complex matrix $m_{\rm D}=\frac{v}{\sqrt{2}}V^{\ell\dag}_LY^\nu$. 
%\\ \\ \\
The lagrangian terms explicitly displayed in Eq.~(\ref{Yukawamass}) incorporate four sorts of Yukawa constants, each of which defines a $3\times3$ matrix: $m_{\rm D}$, $M$, $\mu_S$, and $\mu_R$. In particular, we assume that the matrix $M$ is connected to the stage of spontaneous symmetry breaking taking place at the high-energy scale $\Lambda$. Further, $M\sim\Lambda$ is reasonably assumed. 
The matrix $m_{\rm D}$, on the other hand, is assumed to relate to the Higgs potential vacuum expectation value as $m_{\rm D}\sim v$.
%Moreover, the $3\times3$ matrices $\mu_S=\frac{v_\sigma}{\sqrt{2}}\lambda^{(S)}$ and $\mu_R=\frac{v_\sigma}{\sqrt{2}}\lambda^{(N)}$ emerge as well. 
%\\ \\ \\
The terms involving the $3\times3$ matrices $\mu_S$ and $\mu_R$ violate invariance under the global group ${\rm U}(1)$ and therefore spoil lepton-number conservation. Note that both these matrices are symmetric.
We assume that non-preservation of lepton number is linked to some energy scale $v_\sigma$, and that, in this context, $\mu_S\sim v_\sigma$ and $\mu_R\sim v_\sigma$. For instance, this could be the case of the spontaneous symmetry breaking of the global group ${\rm U}(1)$ discussed in Ref.~\cite{CMP}, from which, among other things, the massless Goldstone boson dubbed ``the majoron'' emerges. Keep in mind that, in general, $\mu_S\ne\mu_R$. Assuming, on the grounds of naturalness~\cite{tHooftnaturalness}, that these matrices are small, which can be achieved if the condition $v_\sigma\ll v$ is fulfilled, ${\rm U}(1)$ global symmetry is only slightly violated. We follow such an assumption from here on. 
%Since these terms are the only ones which do not abide by this global symmetry, notice that the limit in which $\mu_S\to0$ and $\mu_R\to0$ yields a restoration of ${\rm U}(1)$ invariance, thus meaning that these matrices are protected, by ${\rm U}(1)$ symmetry, from receiving large quantum corrections~\cite{}. 
Notice that, from the assumed origin for the $M$ matrix, we have the inverse-seesaw hierarchy condition 
\begin{equation}
v_\sigma\ll v\ll\Lambda,
\label{hierarchy}
\end{equation}
among the three involved energy scales.
%The lepton-number assignments for the fields $S_{j,L}$, $\nu_{j,R}$, and $\sigma$ guarantee invariance of ${\cal L}_{\rm Y}$ with respect to the global group ${\rm U}(1)$. 
The Yukawa sector can be rearranged as
\begin{equation}
{\cal L}_{\rm Y}=\frac{1}{2}
\left(
\begin{array}{ccc}
\overline{\nu_L} & \overline{\nu_R^{\rm c}} & \overline{S_L}
\end{array}
\right)
{\cal M}_\nu
\left(
\begin{array}{c}
\nu_L^{\rm c}
\vspace{0.1cm}
\\
\nu_R
\vspace{0.1cm}
\\
S_L^{\rm c}
\end{array}
\right)
%\nonumber \\ && \hspace{0.8cm}
+{\rm H.c.}+\cdots,
\end{equation}
where ${\cal M}_\nu$ is a $9\times9$ symmetric matrix given, in terms of the $3\times3$ block matrices $m_{\rm D}$, $M$, $\mu_R$, and $\mu_S$, as
\begin{equation}
{\cal M}_\nu=
\left(
\begin{array}{ccc}
0 & m_{\rm D} & 0
\vspace{0.1cm}
\\
m_{\rm D}^{\rm T} & \mu_R & M^{\rm T}
\vspace{0.1cm}
\\
0 & M & \mu_S
\end{array}
\right).
\label{ISSmassmatrix}
\end{equation}
Moreover, the definitions
\begin{equation}
M_{\rm M}=
\left(
\begin{array}{cc}
\mu_R & M^{\rm T}
\vspace{0.1cm}
\\
M & \mu_S
\end{array}
\right),
\end{equation}
%%%%%
\begin{equation}
M_{\rm D}=
\left(
\begin{array}{cc}
m_{\rm D} & 0
\end{array}
\right),
\end{equation}
where the sizes of $M_{\rm M}$ and $M_{\rm D}$ respectively are $6\times6$ and $3\times6$, allow one to give ${\cal M}_\nu$ the form of the neutral-lepton mass matrix that characterizes the type-1 seesaw mechanism. Since ${\cal M}_\nu$ is symmetric, the existence of a unitary-diagonalization matrix $\Omega$ is ensured, with the diagonalization proceeding as~\cite{Takagi}
\begin{equation}
\Omega^{\rm T}{\cal M}_\nu\,\Omega=
\left(
\begin{array}{ccc}
M_n & 0 & 0
\vspace{0.1cm}
\\
0 & M_{N} & 0
\\
0 &0 & M_X
\end{array}
\right).
\label{Omegadiagonalization}
\end{equation}
In this equation, $M_n$ is a $3\times3$ diagonal matrix, whose in-diagonal elements are the light-neutrino masses, which we denote as $m_{n_j}$, with $j=1,2,3$. Furthermore, the matrix $M_{N}$ is $3\times3$ sized and diagonal, with its in-diagonal entries corresponding to the masses $m_{N_j}$ of three heavy neutrinos~%\footnote{Later on, an issue related to wrong signs of mass terms for the eigenvalues $m_{n'_j}$ arises. We discuss this below.}
, so $j=1,2,3$ as well. Finally, the $3\times3$ diagonal matrix $M_X$ comprises, nested within its diagonal, the masses $m_{X_j}$ of three further heavy neutral leptons. This diagonalization procedure entails the following change of basis:
\begin{equation}
\left(
\begin{array}{c}
n_L
\vspace{0.1cm} \\
N_L
\vspace{0.1cm} \\
X_L
\end{array}
\right)
=\Omega^{\rm T}
\left(
\begin{array}{c}
\nu_L
\vspace{0.1cm} \\
\nu_R^{\rm c}
\vspace{0.1cm} \\
S_L
\end{array}
\right),
\hspace{0.3cm}
\left(
\begin{array}{c}
n_R
\vspace{0.1cm} \\
N_R
\vspace{0.1cm} \\
X_R
\end{array}
\right)
=\Omega^\dag
\left(
\begin{array}{c}
\nu_L^{\rm c}
\vspace{0.1cm} \\
\nu_R
\vspace{0.1cm} \\
S_L^{\rm c}
\end{array}
\right).
\end{equation}
Here, $n_L$ and $n_R$ are $3\times1$ matrices for chiral left- and right-handed neutrino fields, respectively. Meanwhile, $N_L$, $N_R$, $X_L$, and $X_R$ are also $3\times1$ sized, comprised by left-handed and right-handed neutral spinor fields. Putting all the pieces together, the Yukawa-sector Lagrangian is written as
\begin{eqnarray}
&&
{\cal L}_{\rm Y}=\sum_{j=1}^3\sum_{k=1}^3\Big(-\frac{1}{2}m_{n_j}\overline{n_{j}}n_{j}
\nonumber \\ && \hspace{0.6cm}
-\frac{1}{2}m_{N_j}\overline{N_j}N_j
%\nonumber \\ && \hspace{0.8cm}
-\frac{1}{2}m_{X_j}\overline{X_j}X_j \Big)+\cdots,
\label{HNLmterms}
\end{eqnarray}
for which the non-chiral fermion fields $n_j=n_{j,L}+n_{j,R}$, $N_j=N_{j,L}+N_{j,R}$, and $X_j=X_{j,L}+X_{j,R}$ have been defined.
\\

The diagonalization matrix $\Omega$ is conveniently expressed as
 \begin{equation}
 \Omega=U\,V,
 \label{Omegadecomposition}
 \end{equation}
with $U$ and $V$ a couple of $9\times9$ unitary matrices. The matrix $U$ is intended to block-diagonalize the mass matrix ${\cal M}_\nu$. This unitary matrix can be expressed in block-matrix form as
\begin{equation}
U=
\left(
\begin{array}{cc}
U_{11} & U_{12}
\vspace{0.1cm} \\
U_{21} & U_{22}
\end{array}
\right),
\label{Ugenpar}
\end{equation}
where $U_{11}$ and $U_{22}$ are square matrices, the former $3\times3$ sized and the latter $6\times6$ sized. Meanwhile, $U_{12}$ is a $3\times6$ matrix and $U_{21}$ is $6\times3$. While the last equation is largely generic, the unitary character of $U$ allows for the following matrix-block parametrization~\cite{KPS,DePi}:
 \begin{equation}
U=
\left(
\begin{array}{cc}
\big( {\bf 1}_3+\xi^*\xi^{\rm T} \big)^{-\frac{1}{2}} & \xi^*\big( {\bf 1}_6+\xi^{\rm T}\xi^* \big)^{-\frac{1}{2}}
\vspace{0.1cm} \\
-\xi^{\rm T}\big( {\bf 1}_3+\xi^*\xi^{\rm T} \big)^{-\frac{1}{2}} & \big( {\bf 1}_6+\xi^{\rm T}\xi^* \big)^{-\frac{1}{2}}
\end{array}
\right),
\label{Ublockmatrixpar}
\end{equation}
where ${\bf 1}_3$ and ${\bf 1}_6$ stand for the $3\times3$ and the $6\times6$ identity matrices, respectively. Further, $\xi$ is a $3\times6$ matrix fulfilling $M_{\rm D}-\xi M_{\rm D}^{\rm T}\xi^*-\xi M_{\rm, M}=0$. If the entries of $\xi$ are small, this condition yields the expression 
\begin{equation}
	\xi=M_{\rm D}M_{\rm M}^{-1},
	\label{xidef}
\end{equation}
corresponding to a large suppression provided by $M_{\rm M}^{-1}$. From here on, we assume Eq.~(\ref{xidef}) to hold. Then, the block parametrization for $U$, Eq.~(\ref{Ublockmatrixpar}), can be approximated as
\begin{widetext}
\begin{equation}
U\approx
\left(
\begin{array}{cc}
{\bf 1}_3-\frac{1}{2}M_{\rm D}^*\big(M_{\rm M}^{-1}\big)^*M_{\rm M}^{-1}M_{\rm D}^{\rm T} & M_{\rm D}^*\big( M_{\rm M}^{-1} \big)^*
\vspace{0.1cm} \\
-M_{\rm M}^{-1}M_{\rm D}^{\rm T} & {\bf 1}_6-\frac{1}{2}M_{\rm M}^{-1}M_{\rm D}^{\rm T}M_{\rm D}^*\big(M_{\rm M}^{-1} \big)^*
\end{array}
\right).
\label{Ublocks}
\end{equation}
\end{widetext}
The afore-announced block-matrix diagonalization, driven by $U$, goes as follows:
\begin{equation}
U^{\rm T}{\cal M}_\nu U=
\left(
\begin{array}{cc}
M_{\rm light} & 0
\vspace{0.1cm}
\\
0 & M_{\rm heavy}
\end{array}
\right),
\end{equation}
where, the matrix $M_{\rm light}=-M_{\rm D}M_{\rm M}^{-1}M_{\rm D}^{\rm T}+{\cal O}\big( (M_{\rm M}^{-1})^3 \big)$ is $3\times3$ sized, whereas $M_{\rm heavy}=M_{\rm M}+{\cal O}\big( M_{\rm M}^{-1} \big)$ is a $6\times6$ matrix. The matrix $V$, introduced in Eq.~(\ref{Omegadecomposition}), is written as
\begin{equation}
V=
\left(
\begin{array}{cc}
\hat{V} & 0
\vspace{0.1cm}
\\
0 & \tilde{V}
\end{array}
\right),
\label{metadiagonalization}
\end{equation}
with the $3\times3$ matrix $\hat{V}$ and the $6\times6$ matrix $\tilde{V}$ respectively diagonalizing $M_{\rm light}$ and $M_{\rm heavy}$ as
\begin{equation}
\hat{V}^{\rm T}M_{\rm light}\, \hat{V}=M_n,
\end{equation}
%%%%%%%%%%
\begin{equation}
\tilde{V}^{\rm T}M_{\rm heavy}\tilde{V}=
\left(
\begin{array}{cc}
	M_{N} & 0
	\vspace{0.1cm}
	\\
	0 & M_X
\end{array}
\right),
\label{Mhdiagonalization}
\end{equation}
in accordance with Eq.~(\ref{Omegadiagonalization}). The inverse of $M_{\rm M}$ turns out to be~\cite{DePi}
\begin{widetext}
\begin{equation}
M_{\rm M}^{-1}=
\left(
\begin{array}{cc}
\big( \mu_R-M^{\rm T}\mu_S^{-1}M \big)^{-1} & -\big( \mu_R-M^{\rm T}\mu_S^{-1}M \big)^{-1}M^{\rm T}\mu_S^{-1}
\vspace{0.1cm}
\\
-\big( \mu_S-M\mu_R^{-1}M^{\rm T} \big)^{-1}M\,\mu_R^{-1} & \big( \mu_S-M\mu_R^{-1}M^{\rm T} \big)^{-1}
\end{array}
\right),
\label{MMinvexplicit}
\end{equation}
\end{widetext}
from which the inverse-seesaw light-neutrino mass formula,
\begin{equation}
M_n\approx \hat{V}^{\rm T}m_{\rm D}M^{-1}\mu_S\,(M^{\rm T})^{-1}m_{\rm D}^{\rm T}\hat{V},
\label{ISSformula}
\end{equation}
follows. According to Eq.~(\ref{ISSformula}), the tinniness characterizing the masses of light neutrinos does not rely only on the suppression introduced by $M^{-1}$, as in the canonical seesaw mechanism, because the smallness of $\mu_S$ establishes a further suppression, thus reducing the stress on the high-energy scale $\Lambda$, associated to $M$, which in this manner evades huge values. 
\\

As shown in Eq.~(\ref{Mhdiagonalization}), the unitary matrix $\tilde{V}$, which diagonalizes the heavy-neutral-lepton mass matrix $M_{\rm heavy}$, yields the diagonal mass matrices $M_{N}$ and $M_X$, which correspond to the sets of heavy-neutral leptons $\{ N_1, N_2, N_3 \}$ and $\{ X_1, X_2, X_3 \}$, respectively. In pursuit of some insight regarding the mass spectra for these fields, let us assume that the matrices $M$, $\mu_R$, and $\mu_S$ are diagonal, in which case $\tilde{V}$ can be written as
\begin{equation}
\tilde{V}=\frac{1}{\sqrt{2}}
\left(
\begin{array}{rcr}
i ({\bf 1}_3+\frac{1}{4}\chi) && {\bf 1}_3-\frac{1}{4}\chi
\vspace{0.2cm} \\ 
-i({\bf 1}_3-\frac{1}{4}\chi) && {\bf 1}_3+\frac{1}{4}\chi
\end{array}
\right),
\label{Vhpar}
\end{equation}
where $\chi=M^{-1}\big( \mu_S-\mu_R \big)$ and $i$ is de imaginary unity. Thus, from Eqs.~(\ref{Mhdiagonalization}) and (\ref{Vhpar}), the diagonal matrices
\begin{equation}
M_{N}=M-\frac{\mu_S}{2}-\frac{\mu_R}{2},
\label{MNmass}
\end{equation}
%%%%%
\begin{equation}
M_X=M+\frac{\mu_S}{2}+\frac{\mu_R}{2},
\end{equation}
emerge, which then lead us to the neutral-lepton mass terms displayed in Eq.~(\ref{HNLmterms}). %In view of the hierarchy of energy scales shown in Eq.~(\ref{hierarchy}), we notice that the mass eigenvalues lying in $M_{n'}$, Eq.~(\ref{MNmass}), are negative, that is, the corresponding heavy-neutral-lepton mass terms bear wrong signs: $+\frac{1}{2}| m_{n'_j} |\overline{n'_j}n'_j$, for $j=1,2,3$. To amend this, we define the fields
%\begin{equation}
%N_j=\gamma_5 n'_j,
%\label{gamma5transformation}
%\end{equation}
%thus getting proper mass terms $-\frac{1}{2}m_{N_j}\overline{N_j}N_j$, with $m_{N_j}=|m_{n'_j}|$. Then the diagonal mass matrix
%\begin{equation}
%M_N=M-\frac{\mu_S}{2}-\frac{\mu_R}{2}
%\label{Nnumassmatrix}
%\end{equation}
%is defined to replace $M_{n'}$, Eq.~(\ref{MNmass}).
Now consider the difference $M_X-M_N=\mu_S+\mu_R$, according to which masses $m_{N_j}$ and $m_{X_j}$, all of them large as dictated by the hierarchy condition shown in Eq.~(\ref{hierarchy}), are very similar to each other. Thus the complete heavy-neutral-lepton mass spectrum is quasi-degenerate for each pair of fields $N_j$ and $X_j$, with $j=1,2,3$.
\\

To recapitulate, after the spontaneous symmetry breaking of the Standard-Model electroweak gauge group ${\rm SU}(2)_L\otimes{\rm U}(1)_Y$ into the electromagnetic group ${\rm U}(1)_e$, at $v$, and the ulterior breaking of the global ${\rm U}(1)$ symmetry at $v_\sigma$, the neutral-lepton field content comprehends three light-neutrino fields, denoted by $n_j$, also three heavy-neutrino fields, which we have referred to as $N_j$, and three further heavy-neutral-lepton fields, represented by $X_j$, with $m_{N_j}\approx m_{X_j}$, for each pair labeled by $j=1,2,3$. Let us jointly denote all the heavy-neutral-lepton fields as
\begin{equation}
f_j=
\left\{
\begin{array}{l}
%n_k,\textrm{ if }j=1,2,3,
%\vspace{0.1cm} \\
N_k,\textrm{ if }j=1,2,3,
\vspace{0.1cm} \\
X_k,\textrm{ if }j=4,5,6,
\end{array}
\right.
\end{equation}
where $N_k=N_1,N_2,N_3$ and $X_k=X_1,X_2,X_3$. With this in mind, the charged-currents lagrangian term can be written as
\begin{eqnarray}
&&
{\cal L}_{W\nu l}=\frac{-g}{\sqrt{2}}\sum_{\alpha=e,\mu,\tau}
\Big(
\sum_{j=1}^3
\big(
{\cal B}_{\alpha n_j}W^-_\mu\overline{l_\alpha}\gamma^\mu P_Ln_j+{\rm H.c.}
\big)
\vspace{0.1cm} \nonumber \\ && \hspace{1.2cm}
+\sum_{j=1}^6\big(
{\cal B}_{\alpha f_j}W^-_\mu\overline{l_\alpha}\gamma^\mu P_Lf_j+{\rm H.c.}
\big)
\Big),
\label{Wnul}
\end{eqnarray}
where $g$ is the weak coupling constant. Moreover, the definitions
\begin{equation}
{\cal B}_{\alpha n_j}=\big( U_{11}^* \hat{V}^* \big)_{\alpha j},
\label{Blight}
\end{equation}
%%%%%%%%%%
\begin{equation}
{\cal B}_{\alpha f_j}=\big( U_{12}^* \tilde{V}^* \big)_{\alpha j},
\label{Bheavy}
\end{equation}
have been used. The factors $\big( U_{11} \big)_{\alpha \beta}$ and $\big( U_{12} \big)_{\alpha k}$, in Eqs.~(\ref{Blight}) and (\ref{Bheavy}), are components of the matrices $U_{11}$ and $U_{12}$, which are part of the generic block-matrix form of $U$, given in Eq.~(\ref{Ugenpar}). Moreover, $\hat{V}_{\beta j}$ and $\tilde{V}_{kj}$ are entries of the unitary matrices introduced in Eq.~(\ref{metadiagonalization}). The whole set of quantities ${\cal B}_{\alpha n_j}$ and ${\cal B}_{\alpha f_j}$ constitute the $3\times3$ matrix ${\cal B}_n$ and the $3\times6$ matrix ${\cal B}_f$, respectively, which can be accommodated into the $3\times9$ matrix ${\cal B}=\big( {\cal B}_n\,,\,{\cal B}_f \big)$, with entries
\begin{equation}
{\cal B}_{\alpha j}=
\left\{
\begin{array}{l}
{\cal B}_{\alpha n_k},\textrm{ if }j=1,2,3,
\vspace{0.2cm} \\
{\cal B}_{\alpha f_k},\textrm{ if }j=4,5,6,7,8,9,
\end{array}
\right.
\end{equation}
where $n_k=n_1,n_2,n_3$ and $f_k=f_1,f_2,f_3,f_4,f_5,f_6$, for any fixed $\alpha$. The matrix ${\cal B}$ fulfills a sort of semi-unitarity property, that is
\begin{eqnarray}
\displaystyle
\sum_{j=1}^9{\cal B}_{\alpha j}{\cal B}_{\beta j}^*=\delta_{\alpha\beta},
\label{semiuny1}
\vspace{0.1cm}
\\  \displaystyle
%\end{equation}
%%%%%%%%%%
%\begin{equation}
\sum_{\alpha=e,\mu,\tau}{\cal B}_{\alpha j}^*{\cal B}_{\alpha k}={\cal C}_{jk}.
\label{semiuny2}
\end{eqnarray}
The quantities ${\cal C}_{jk}$, involved in Eq.~(\ref{semiuny2}), form a $9\times9$ matrix, ${\cal C}$, which can be written in block-matrix form as
\begin{equation}
{\cal C}=
\left(
\begin{array}{cc}
{\cal C}_{nn} & {\cal C}_{nf}
\vspace{0.1cm} \\
{\cal C}_{fn} & {\cal C}_{ff}
\end{array}
\right),
\end{equation}
where ${\cal C}_{nn}$ is $3\times3$, the size of ${\cal C}_{ff}$ is $6\times6$, ${\cal C}_{nf}$ is a $3\times6$ matrix, and ${\cal C}_{fn}$ is $6\times3$ sized. Furthermore, the entries of ${\cal C}$ relate to those of these block matrices as
\begin{equation}
{\cal C}_{jk}=
\left\{
\begin{array}{l}
C_{n_ln_i},\textrm{ if }j=1,2,3\textrm{ and }k=1,2,3,
\vspace{0.2cm} \\
C_{n_lf_i},\textrm{ if }j=1,2,3\textrm{ and }k=4,5,6,7,8,9,
\vspace{0.2cm} \\
C_{f_ln_i},\textrm{ if }j=4,5,6,7,8,9\textrm{ and }k=1,2,3,
\vspace{0.2cm} \\
C_{f_lf_i},\textrm{ if }j=4,5,6,7,8,9\textrm{ and }k=4,5,6,7,8,9,
\end{array}
\right.
\end{equation}
where $n_l,n_i=n_1,n_2,n_3$ and $f_l,f_i=f_1,f_2,f_3,f_4,f_5,f_6$. These block matrices are defined as
\begin{equation}
{\cal C}_{n_jn_k}=\big( \Omega_{11}^{\rm T}\Omega_{11}^* \big)_{jk},
\end{equation}
%%%%%%%%%%
\begin{equation}
{\cal C}_{n_jf_k}=\big( \Omega_{11}^{\rm T}\Omega_{12}^* \big)_{jk},
\end{equation}
%%%%%%%%%%
\begin{equation}
{\cal C}_{f_jn_k}=\big( \Omega_{12}^{\rm T}\Omega_{11}^* \big)_{jk},
\end{equation}
%%%%%%%%%%
\begin{equation}
{\cal C}_{f_jf_k}=\big( \Omega_{12}^{\rm T}\Omega_{12}^* \big)_{jk}.
\end{equation}
\\

From the $U$ block parametrization displayed in Eq.~(\ref{Ublocks}), the expression for $M^{-1}_{\rm M}$ as given in Eq.~(\ref{MMinvexplicit}), and the explicit form shown in Eq.~(\ref{Vhpar}) for $\tilde{V}$, the matrices ${\cal B}_n$ and ${\cal B}_f$ can be written as
\begin{equation}
{\cal B}_n=\Big( {\bf 1}_3-\frac{1}{2}m_{\rm D}M^{-2}m_{\rm D}^\dag \Big)\hat{V}^*,
\label{BnISS}
\end{equation}
%%%%%%%%%%
\begin{equation}
{\cal B}_f=
\Big( 
\begin{array}{lcr}
	\frac{i}{\sqrt{2}}m_{\rm D}M^{-1} && \frac{1}{\sqrt{2}}m_{\rm D}M^{-1}
\end{array}
\Big).
\label{BfISS}
\end{equation}
In the so-called ``minimally extended Standard Model''~\cite{GiKi}, characterized by the addition of three right-handed Dirac-neutrino fields, Yukawa terms for right-handed neutrinos give rise to neutrino mass terms. In the neutrino mass-eigenspinor basis, the lepton charged currents, $j^\mu_{{\rm SM},W}=2\,\overline{l_L}\,U_{\rm PMNS}\gamma^\mu\,n_L$, involve lepton-flavor change, driven by the PMNS matrix, $U_{\rm PMNS}$, which is the lepton-sector analogue of the Kobayashi-Maskawa matrix, lying in the quark sector. The neutrino-flavor basis is then defined by the transformation $\nu_L=U_{\rm PMNS}\,n_L$, by means of which the charged currents are simply expressed as $j^\mu_{{\rm SM},W}=2\,\overline{l_L}\gamma^\mu\,\nu_L$. An important feature of $U_{\rm PMNS}$ is its unitary property. A similar discussion can be developed if neutrino masses are rather defined through the inclusion, in the framework of a lepton-number-violating effective Lagrangian for Standard Model, of the Weinberg operator. Getting back to the theoretical framework of the present paper, the charged currents in which light neutrinos participate, displayed in the first line of Eq.~(\ref{Wnul}), carry neutrino mixing characterized not by the $U_{\rm PMNS}$ unitary matrix, but by the $3\times3$ matrix ${\cal B}_n$, which is not unitary. Expressing this matrix as ${\cal B}_n=\big( {\bf 1}_3-\eta \big)\hat{V}^*$, the $3\times3$ matrix
\begin{equation}
\eta=\frac{1}{2}m_{\rm D}M^{-2}m_{\rm D}^\dag
\label{nonunitaritymatrix}
\end{equation}
is understood as an object characterizing non-unitary effects in light-neutrino mixing.% The expression displayed in Eq.~(\ref{BfISS}) for ${\cal B}_f$ comprises two block matrices, both of them given as $\frac{1}{\sqrt{2}}m_{\rm D}M^{-1}$. After the two stages of spontaneous symmetry breaking, at $v$ and at $v_\sigma$, and the diagonalization process, implemented by $\Omega$, the expression for the first block matrix was actually $-\frac{1}{\sqrt{2}}m_{\rm D}M^{-1}$. The reason behind the different sign displayed in Eq.~(\ref{BfISS}) lies in the transformation given in Eq.~(\ref{gamma5transformation}), introduced to correct the wrong signs in heavy-neutrino mass terms, as it produces an extra negative sign in the charged currents in which heavy-neutrino fields $N_j$ participate. Then, such a sign is conveniently absorbed in the definition of ${\cal B}_f$.%
\\

%%%%%%%%%%%%%%%%%%%%
%%%%%%%%%%%%%%%%%%%%
%%%%%%%%%%%%%%%%%%%%
%%%%%%%%%%%%%%%%%%%%
%%%%%%%%%%%%%%%%%%%%

\section{Lepton-flavor change induced by the inverse seesaw at one loop}
\label{analytic}
In this section, we present and discuss our analytical calculation of contributions from virtual light neutrinos neutrinos $n_j$, heavy neutrinos $N_j$, and heavy neutral leptons $X_j$, to the charged-lepton-flavor-violating decay $l_\alpha\to l_\beta\,\gamma$, where the indices $\alpha$ and $\beta$ label charged-lepton flavors, so $\alpha=\mu,\tau$ and $\beta=e,\mu$. In the presence of massive neutrinos, and as long as neutrino mixing takes place, such a process receives contributions from Feynman diagrams since the one-loop order, whatever Standard-Model extension is considered. Note that, in contraposition, this decay process is forbidden in the Standard Model. In general, physical processes which are either not allowed by the Standard Model or suppressed in such a framework bear great relevance, because manifestations from the high-energy description might be more easily detected. 
\\

In the context of the inverse seesaw mechanism, the contributing diagrams emerge at one loop as a result of lepton mixing in the charged currents displayed in Eq.~(\ref{Wnul}), which are characterized by the matrix ${\cal B}=\big( {\cal B}_n,{\cal B}_f \big)$. In general, the set of Feynman rules by which Majorana fermions abide differ from those for fermions described by Dirac fields. Refs.~\cite{DEHK,GlZr} provide detailed discussions on the matter. For instance, the assumption of Majorana fermions opens the possibility of having a larger number of contributing diagrams, as compared to the Dirac case. However, by following the Wick's theorem~\cite{Wick}, we found no extra diagrams. Moreover, all the determined contributing diagrams were found not to distinguish among the Majorana and Dirac cases. 
\\

While gauge invariance is an essential element in the construction of field theories, as it provides a criterion to build Lagrangian terms, the quantization of gauge theories imperiously requires the elimination of this symmetry, which is achieved by fixing the gauge. Then, gauge freedom ensures that any physical quantity must be gauge independent. The linear and the non-linear gauge-fixing approaches~\cite{FLS,Shore,EiWu,DDW,MeTo}, where the choice of the gauge is parametrized by some gauge-fixing parameter, $\xi$, are well-known gauge-fixing schemes. Another, often considered, gauge choice is the ``unitary gauge'', distinguished by the absence of pseudo-Goldstone bosons, which are born in the process of spontaneous symmetry breaking but which lack physical degrees of freedom. In particular, the gauge choice must be innocuous to the decay amplitude we are calculating. Such an amplitude is given by $i{\cal M}=\overline{u_\beta}\,\Gamma_\mu^{\beta\alpha} u_\alpha\,\varepsilon^{\mu*}$, where $u_\alpha$ and $u_\beta$ are momentum-space spinors, whereas $\varepsilon^\mu$ is the polarization vector associated to the electromagnetic field. In general, the vertex function $\Gamma_\mu^{\beta\alpha}$ can be expressed as
\begin{widetext}
\begin{eqnarray}
	&&
\Gamma_\mu^{\beta\alpha}=
	\hspace{0.2cm}
\sum_{j=1}^3\sum_{\psi=n,N,X}
\Bigg(
\begin{gathered}
	\hspace{0.2cm}
	\vspace{-0.2cm}
	\includegraphics[width=2.3cm]{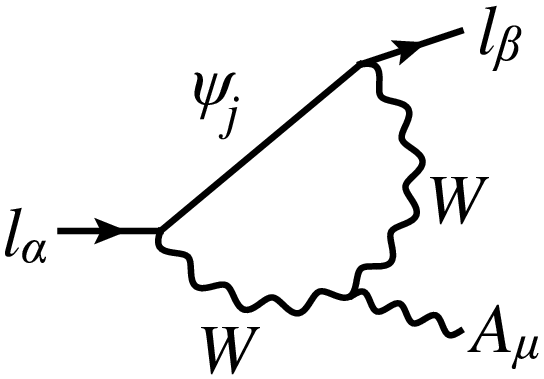}
\end{gathered}
+
\begin{gathered}
	\hspace{0.2cm}
	\vspace{-0.2cm}
	\includegraphics[width=2.3cm]{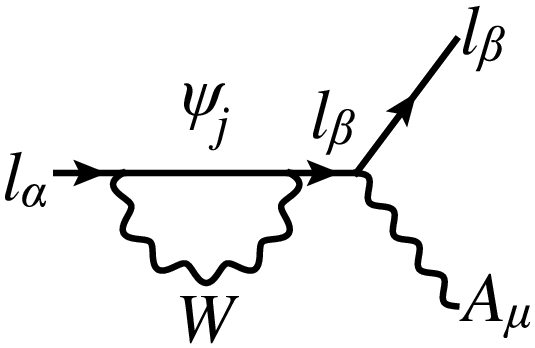}
\end{gathered}
	+
\begin{gathered}
	\hspace{0.2cm}
	\vspace{-0.2cm}
	\includegraphics[width=2.3cm]{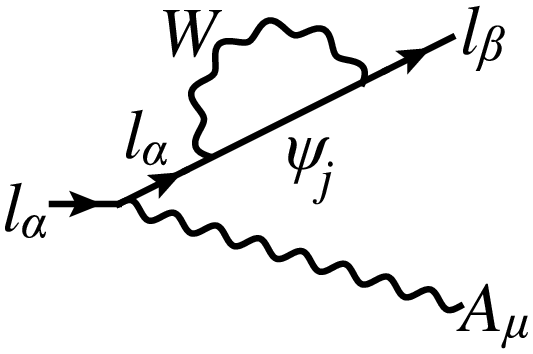}
\end{gathered}
\nonumber \\ \nonumber \\ \nonumber \\ \nonumber && \hspace{3cm}
+
\begin{gathered}
	\hspace{0.2cm}
	\vspace{-0.2cm}
	\includegraphics[width=2.3cm]{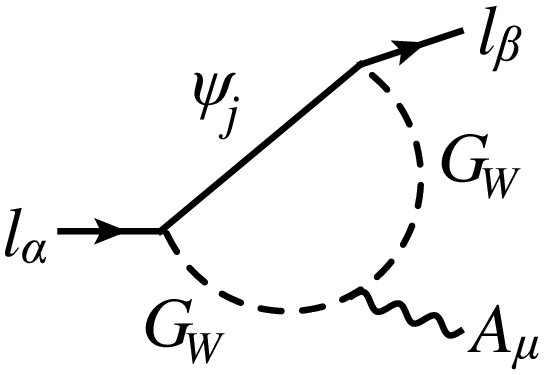}
\end{gathered}
+
\begin{gathered}
	\hspace{0.2cm}
	\vspace{-0.2cm}
	\includegraphics[width=2.3cm]{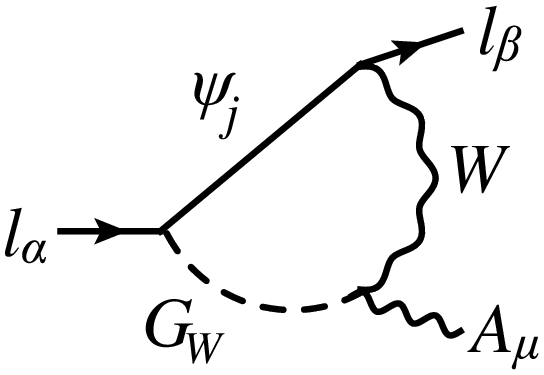}
\end{gathered}
	+
\begin{gathered}
	\hspace{0.2cm}
	\vspace{-0.2cm}
	\includegraphics[width=2.3cm]{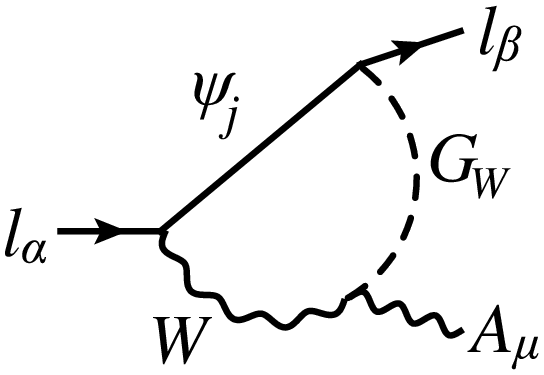}
\end{gathered}
\nonumber \\ \nonumber \\ \nonumber \\ && \hspace{3cm}
+
\begin{gathered}
	\hspace{0.2cm}
	\vspace{-0.2cm}
	\includegraphics[width=2.3cm]{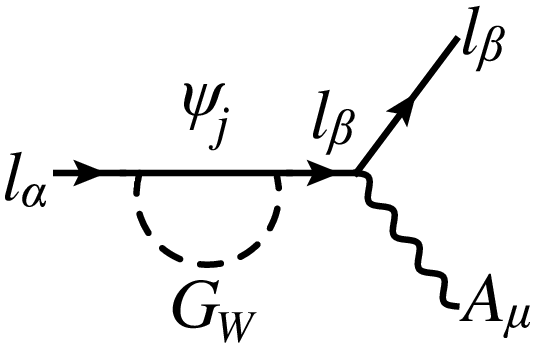}
\end{gathered}
+
\begin{gathered}
	\hspace{0.2cm}
	\vspace{-0.2cm}
	\includegraphics[width=2.3cm]{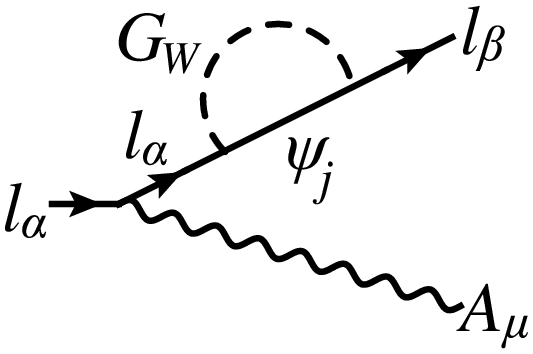}
\end{gathered}
\Bigg),
\label{diagramaticamp}
\end{eqnarray}
\end{widetext} 
where $l_\alpha$ and $l_\beta$ stand for $\alpha$- and $\beta$-flavored charged leptons, and $W$ represents a $W$ boson, with $G_W$ its associated pseudo-Goldstone boson. Also, neutral leptons, light ones and heavy ones as well, are generically denoted as $\psi_j=n_j,N_j,X_j$. We carried out the calculation of the amplitude for $l_\alpha\to\,l_\beta\,\gamma$ using the non-linear gauge approach and the unitary gauge as well. As far as the non-linear gauge-fixing approach is concerned, use has been made of the ${\rm U}(1)_e$-covariant gauge-fixing functions 
\begin{equation}
f^{\pm}_{\rm GF}=D_\mu W^{\pm \mu}-i\xi m_WG_W^{\pm},
\label{GFfunctions}
\end{equation}
thoroughly discussed in Ref.~\cite{MeTo}. Here, $D_\mu$ is the electromagnetic covariant derivative. An outcome of the implementation of Eq.~(\ref{GFfunctions}) to fix the gauge is the modification of some gauge-boson vertices, such as $WW\gamma$, which in this manner becomes gauge dependent. The resulting $WW\gamma$ vertex reads
\begin{eqnarray}
&&
\begin{gathered}
\vspace{-0.2cm}
\includegraphics[width=2.3cm]{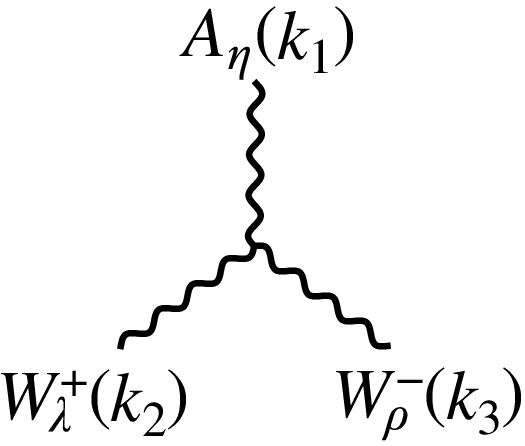}
\end{gathered}
=
ie
\Big(
\big( k_3-k_2 \big)_\eta g_{\rho\lambda}
\nonumber \\ && \hspace{2.8cm}
+\Big( k_1-k_3-\frac{1}{\xi}k_2 \Big)_\lambda g_{\rho\eta}
\nonumber \\ 
\nonumber \\
\nonumber \\ && \hspace{2.5cm}
+\Big( k_2-k_1+\frac{1}{\xi}k_3 \Big)_\rho g_{\lambda\eta}
\Big).
\nonumber \\
\label{nlWWgamma}
\end{eqnarray}
This change is compensated by the elimination of the vertices $WG_W\gamma$, thus meaning that the number of Feynman diagrams to calculate, for some given process, is reduced. In our case, since these vertices are pieces of the 5th and 6th diagrams in Eq.~(\ref{diagramaticamp}), such contributions are absent. To execute our calculation, we have used the Feynman-t' Hooft gauge, that is, we take value $\xi=1$, for the gauge-fixing parameter. Also, the following $G_W\nu l$ Lagrangian has been used:
\begin{eqnarray}
&&
{\cal L}_{G_W\nu l}=\frac{-g}{\sqrt{2} m_W}\sum_{\alpha=e,\mu,\tau}
\nonumber \\ &&\times
\Big(
\sum_{j=1}^3
\big(
{\cal B}_{\alpha n_j}G^-_W\overline{l_\alpha}\left( m_\alpha P_L - m_{n_j} P_R \right) n_j+{\rm H.c.}
\big) \vspace{0.1cm} \nonumber \\ && %\hspace{1.2cm}
+\sum_{j=1}^6\big(
{\cal B}_{\alpha f_j}G^-_W\overline{l_\alpha}\left( m_\alpha P_L - m_{f_j} P_R \right)f_j+{\rm H.c.}
\big)
\Big),
\nonumber \\
\label{GWnul}
\end{eqnarray}
where $m_\alpha$ is the mass of the charged lepton $l_\alpha$, whereas $P_L$ and $P_R$ are the chiral projection matrices. On the other hand, in the case of the unitary gauge, by definition, no contributing Feynman diagrams featuring pseudo-Goldstone-boson lines are to be taken into account, in contraposition to what happens in other gauges. Therefore, in this gauge only the first three diagrams contribute. Let us remark that, as expected on the grounds of consistency, we have found that the calculation of ${\cal M}$ yields the same results in both gauges, which shows that our result is gauge independent.
\\

The presence of loop-momentum integrals opens the possibility of having ultraviolet-divergent contributions. About this, note that the superficial degrees of divergence of the diagrams shown in Eq.~(\ref{diagramaticamp}) are 0, 1, and 1, respectively, so these pieces might bear ultraviolet divergences growing as fast as linearly. To deal with this, a regularization method must be implemented. Among the different options, we followed the dimensional-regularization approach~\cite{BoGi,tHVe}. This regularization method has the advantage of preserving gauge symmetry, ensured by fulfillment of Ward identities~\cite{Ward}, and is also well suited for its implementation through software tools. In the dimensional-regularization method, the dimension of spacetime is assumed to be $D$, with 1 time-like dimension and $D-1$ space-like dimensions. Then, loop integrals are modified as $\int\frac{d^4k}{(2\pi)^4}F(k)\to\mu_{\rm R}^{4-D}\int\frac{d^Dk}{(2\pi)^D}F(k)$, where $\mu_{\rm R}$ is the renormalization scale, which has units of mass and whose task is to preserve the units of loop integrals, whereas $F(k)$ represents some function of the loop 4-momentum $k$. An analytic continuation of the $D$-dimensional loop integrals is defined by assuming the dimension $D$ to be a complex quantity, with $D\to4$.
\\

\begin{figure}[ht]
\center
\includegraphics[width=7cm]{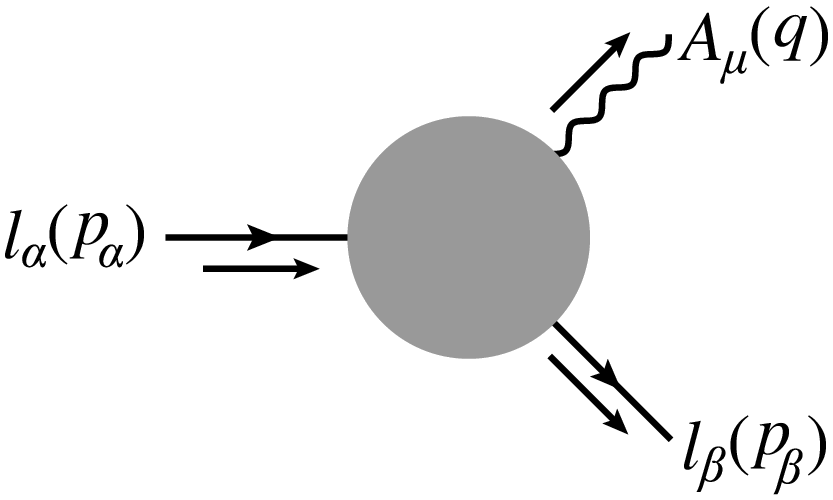}
\caption{\label{llAdecay} Conventions for momenta used to carry out the calculation of the $l_\alpha\to\,l_\beta\gamma$ amplitude.}
\end{figure}

To perform the analytic calculation of the amplitudes corresponding to the contributing diagrams of Eq.~(\ref{diagramaticamp}), use has been made of the software packages \textsc{FeynCalc}~\cite{SMO1,SMO2,MBD} and \textsc{Package-X}~\cite{Patel}, implemented through \textsc{Mathematica}, by Wolfram. With these software tools, calculations have been carried out by following the tensor-reduction method~\cite{PaVe,DeSt}. Furthermore, the momenta conventions displayed in Fig.~\ref{llAdecay}
have been used to perform the calculation, where $q=p_\alpha-p_\beta$, due to 4-momentum conservation. The expression we found for the vertex function $\Gamma_\mu^{\beta\alpha}$ has the gauge-invariant and Lorentz-covariant structure
\begin{equation}
\Gamma_\mu^{\beta\alpha}=\mu^{\beta\alpha}\,\sigma_{\mu\nu}q^\nu+d^{\beta\alpha}\,\sigma_{\mu\nu} q^\nu\gamma_5,
\end{equation}
where $\mu^{\beta\alpha}$ and $d^{\beta\alpha}$ are the transition magnetic form factor and the transition electric form factor, respectively~\cite{NPR,BGS}. The corresponding branching ratio is then given by
\begin{equation}
{\rm Br}(l_\alpha\to\,l_\beta\gamma)=\frac{\big( m_\alpha^2-m_\beta^2 \big)^3}{8\pi m_\alpha^3\Gamma_{\rm tot.}}\Big( | \mu^{\beta\alpha} |^2+| d^{\beta\alpha} |^2 \Big),
\label{exactBr}
\end{equation}
where $m_\alpha$ and $m_\beta$ respectively denote the masses of the charged-leptons $l_\alpha$ and $l_\beta$, whereas $\Gamma_{\rm tot.}$ is the total decay rate for $l_\alpha$.
\\

The transition magnetic and electric form factors, $\mu^{\beta\alpha}$ and $d^{\beta\alpha}$, can be written as
\begin{equation}
\mu^{\beta\alpha}=\sum_{j=1}^3{\cal B}_{\beta n_j}{\cal B}_{\alpha n_j}^*\mu^{\beta\alpha}_{n_j}+\sum_{j=1}^6{\cal B}_{\beta f_j}{\cal B}_{\alpha f_j}^*\mu^{\beta\alpha}_{f_j},
\label{mubetaalpha}
\end{equation}
%%%%%
\begin{equation}
d^{\beta\alpha}=\sum_{j=1}^3{\cal B}_{\beta n_j}{\cal B}_{\alpha n_j}^*d^{\beta\alpha}_{n_j}+\sum_{j=1}^6{\cal B}_{\beta f_j}{\cal B}_{\alpha f_j}^*d^{\beta\alpha}_{f_j}.
\label{dbetaalpha}
\end{equation}
The $\mu^{\beta\alpha}_{n_j}$ and $d^{\beta\alpha}_{n_j}$ factors, corresponding to the sum of contributing Feynman diagrams which exclusively involve the $j$-th virtual light neutrino $n_j$, are functions of the mass $m_{n_j}$, whereas $\mu^{\beta\alpha}_{f_j}$ and $d^{\beta\alpha}_{f_j}$, associated to the sum of diagrams in which only the $j$-th virtual heavy-neutral lepton $f_j$ participates, is either $m_{N_j}$ or $m_{X_j}$ dependent. All the aforementioned form-factor contributions depend on the masses $m_\alpha$ and $m_\beta$, of the external leptons, while they also depend on the $W$-boson mass, $m_W$. Moreover, since use has been made of the tensor-reduction method, the resulting expressions for all these contributing form factors are given in terms of Passarino-Veltman scalar functions~\cite{PaVe}. More precisely, 2-point and 3-point Passarino-Veltman scalar functions, respectively defined as
\begin{widetext}
\begin{equation}
B_0(p_1^2,m_0^2,m_1^2)=\frac{(2\pi\mu_{\rm R})^{4-D}}{i\pi^2}\int d^Dk\frac{1}{\big( k^2-m_0^2 \big)\big( (k+p_1)^2-m_1^2 \big)},
\end{equation}
%%%%%%%%%%
\begin{eqnarray}
&&
C_0(p_1^2,(p_1-p_2)^2,p_2^2,m_0^2,m_1^2,m_2^2)
%\nonumber \\ && \hspace{0.9cm}
=\frac{(2\pi\mu_{\rm R})^{4-D}}{i\pi^2}\int d^Dk\frac{1}{\big( k^2-m_0^2 \big)\big( (k+p_1)^2-m_1^2 \big)\big( (k+p_2)^2-m_2^2 \big)},
\label{PaVeC0}
\end{eqnarray}
\end{widetext}
are involved.
%The explicit expressions of the electromagnetic-form-factor partial contributions $\mu^{\beta\alpha}_{n_j}$, $\mu^{\beta\alpha}_{f_j}$, $d^{\beta\alpha}_{n_j}$, and $d^{\beta\alpha}_{f_j}$ are provided in the Appendix.
The scalar functions $B_0$ are the sources of ultraviolet divergences, whereas 3-point functions $C_0$ are finite in this sense. 
%A latent drawback of the unitary gauge, regarding the ultraviolet behavior of the amplitude $\Gamma_\mu^{\beta\alpha}$, is that the growth of ultraviolet divergences might be worsened because, by this election, the gauge-boson propagators increase the superficial degree of divergence of the diagrams and, thus, has the potential of complicating the elimination of ultraviolet divergences. 
Let us point out that any factor $\mu^{\beta\alpha}_{n_j}$, $\mu^{\beta\alpha}_{f_j}$, $d^{\beta\alpha}_{n_j}$, or $d^{\beta\alpha}_{f_j}$ has the following structure: $\sum_kB_0^{(k)}f^{(k)}+\sum_nC_0^{(n)}g^{(n)}+h$, where $f^{(k)}$ is a mass-dependent function accompanied by some 2-point scalar function, $B_0^{(k)}$, whereas $g^{(n)}$ is another function depending on masses, which appears multiplied by a 3-point scalar function, here referred to as $C_0^{(n)}$. The sums $\sum_k$ and $\sum_n$ run over all the 2- and 3-point Passarino-Veltman scalar functions featured by the partial contribution under consideration. Finally, $h$ is another mass-dependent function, which is not multiplied by any loop scalar function. We have been able to verify that the first term, $\sum_kB_0^{(k)}f^{(k)}$, is written as a sum made exclusively of terms of the form $\big( B_0^{(r)}-B_0^{(s)} \big)w^{(r,s)}$, with $w^{(r,s)}$ some combination of the aforementioned mass-dependent functions $f^{(k)}$. In other words, any partial form-factor contribution $\mu^{\beta\alpha}_{n_j}$, $\mu^{\beta\alpha}_{f_j}$, $d^{\beta\alpha}_{n_j}$, or $d^{\beta\alpha}_{f_j}$ can be written in such a way that the presence of 2-point functions exclusively occurs as differences among pairs of them. Keep in mind that all the $B_0$ functions, no matter what their arguments are, share the same divergent part, that is, $B_0^{(k)}=\Delta_{\rm div.}+\eta^{(k)}_{\rm fin.}$ for all $k$, where $\Delta_{\rm div.}$ gathers all the divergent contribution, whereas $\eta^{(k)}_{\rm fin.}$ represents the non-divergent part, which is determined by the specific arguments of the 2-point function under consideration. Then, differences $B_0^{(r)}-B_0^{(s)}$ are ultraviolet finite, thus implying that $\mu^{\beta\alpha}_{n_j}$, $\mu^{\beta\alpha}_{f_j}$, $d^{\beta\alpha}_{n_j}$, or $d^{\beta\alpha}_{f_j}$ are free of ultraviolet divergences. 
%In conclusion, despite our gauge choice the electromagnetic form factors turn out to be ultraviolet finite. 
\\

While the matrix ${\cal B}$ is not unitary, Eq.~(\ref{semiuny1}) allows for a sort of Glashow-Iliopoulos-Maiani mechanism to operate~\cite{GIM}. An adequate implementation of this mechanism, during numerical evaluation, is imperative in order to avoid inaccurately large contributions. By usage of Eq.~(\ref{semiuny1}), with the flavor-change assumption $\alpha\ne\beta$, the transition electromagnetic moments $\mu^{\beta\alpha}$ and $d^{\beta\alpha}$ can be conveniently written as
\begin{eqnarray}
&&
\mu^{\beta\alpha}=\sum_{j=1}^3{\cal B}_{\beta n_j}{\cal B}_{\alpha n_j}^*\big( \mu_{n_j}^{\beta\alpha}-\mu_{f_6}^{\beta\alpha} \big)
\nonumber \\ && \hspace{0.8cm}
+\sum_{j=1}^5{\cal B}_{\beta f_j}{\cal B}_{\alpha f_j}^*\big( \mu_{f_j}^{\beta\alpha}-\mu_{f_6}^{\beta\alpha} \big),
\label{muGIM}
\end{eqnarray}
%%%%%%%%%%
\begin{eqnarray}
&&
d^{\beta\alpha}=\sum_{j=1}^3{\cal B}_{\beta n_j}{\cal B}_{\alpha n_j}^*\big( d_{n_j}^{\beta\alpha}-d_{f_6}^{\beta\alpha} \big)
\nonumber \\ && \hspace{0.8cm}
+\sum_{j=1}^5{\cal B}_{\beta f_j}{\cal B}_{\alpha f_j}^*\big( d_{f_j}^{\beta\alpha}-d_{f_6}^{\beta\alpha} \big).
\label{dGIM}
\end{eqnarray}
Now notice that all contributions $\mu^{\beta\alpha}_{n_1}$, $\mu^{\beta\alpha}_{n_2}$, $\mu^{\beta\alpha}_{n_3}$, $\mu^{\beta\alpha}_{f_1}$, $\mu^{\beta\alpha}_{f_2}$, $\mu^{\beta\alpha}_{f_3}$, $\mu^{\beta\alpha}_{f_4}$, $\mu^{\beta\alpha}_{f_5}$, $\mu^{\beta\alpha}_{f_6}$ come from Feynman diagrams sharing the very same structure, only distinguished among each other by the virtual neutral lepton flowing through the loop, which can be appreciated from the diagrammatic expression displayed in Eq.~(\ref{diagramaticamp}). For this reason, these partial contributions to $\mu^{\beta\alpha}$ differ of each other only by their neutral-lepton-mass dependence. For instance, if the change $m_{n_3}\to m_{f_5}$ is implemented in $\mu_{n_3}^{\beta\alpha}$, the resulting expression is the one for $\mu_{f_5}^{\beta\alpha}$. The same argumentation applies for the electric-dipole partial contributions $d^{\beta\alpha}_{n_1}$, $d^{\beta\alpha}_{n_2}$, $d^{\beta\alpha}_{n_3}$, $d^{\beta\alpha}_{f_1}$, $d^{\beta\alpha}_{f_2}$, $d^{\beta\alpha}_{f_3}$, $d^{\beta\alpha}_{f_4}$, $d^{\beta\alpha}_{f_5}$, $d^{\beta\alpha}_{f_6}$. Then note that terms which do not depend on heavy-neutral-lepton mass vanish from the differences $\big( \mu_{n_j}^{\beta\alpha}-\mu_{f_6}^{\beta\alpha} \big)$, $\big( \mu_{f_j}^{\beta\alpha}-\mu_{f_6}^{\beta\alpha} \big)$, $\big( d_{n_j}^{\beta\alpha}-d_{f_6}^{\beta\alpha} \big)$, and $\big( d_{f_j}^{\beta\alpha}-d_{f_6}^{\beta\alpha} \big)$, in Eqs.~(\ref{muGIM}) and (\ref{dGIM}). Moreover, notice that further cancellations from some of these differences take place, thus yielding a delicate balance in which a fine suppression of contributions happens.
\\

%%%%%%%%%%%%%%%%%%%%
%%%%%%%%%%%%%%%%%%%%
%%%%%%%%%%%%%%%%%%%%
%%%%%%%%%%%%%%%%%%%%
%%%%%%%%%%%%%%%%%%%%

\section{Estimations and discussion of the contributions}
\label{numbers}
Now we turn to our numerical estimations of contributions. Charged-lepton-flavor-violating decays $l_\alpha\to\,l_\beta\,\gamma$ serve as means to search for new-physics traces. On the other hand, in the presence of nonzero neutrino masses and lepton mixing, these decays can be generated since the one-loop level. Up to these days, such decay processes have never been observed, while stringent bounds are available~\cite{PDG}. In Ref.~\cite{MEGlfvbound}, the MEG Collaboration reported their results on a study of the decay $\mu^+\to e^+\gamma$, from which the upper limit ${\rm Br}\big( \mu^+\to e^+\gamma \big)_{\rm MEG}<4.2\times10^{-13}$, at the $90\%\,{\rm C.L.}$, was set on the branching ratio for this decay. Moreover, according to Ref.~\cite{MEG2lfvestimation}, the upcoming MEG II detector will be able to search for $\mu^+\to e^+\gamma$ with an improved sensitivity of $6\times10^{-14}$. A search for the tau lepton-flavor-violating decays $\tau\to e\gamma$ and $\tau\to\mu\gamma$ was reported in Ref.~\cite{BaBarlfvbound}, by the BaBar Collaboration, where the bounds ${\rm Br}\big( \tau\to e\gamma \big)_{\rm BaBar}<3.3\times10^{-8}$ and ${\rm Br}\big( \tau\to\mu\gamma \big)_{\rm BaBar}<4.4\times10^{-8}$ were derived, both at the $90\%\,{\rm C.L.}$ A more recent analysis on these tau decays has been carried out by the Belle Collaboration, which in Ref.~\cite{Bellelfvbound} presented the limits ${\rm Br}\big( \tau\to e\gamma \big)_{\rm Belle}<5.6\times10^{-8}$ and ${\rm Br}\big( \tau\to\mu\gamma \big)_{\rm Belle}<4.2\times10^{-8}$ at the $90\%\,{\rm C.L.}$ Note that the Belle II Collaboration has projected an increased sensitivity to these tau decays of order $10^{-9}$, namely, bounds as stringent as ${\rm Br}\big( \tau\to e\gamma \big)_{\rm Belle\,II}<9.0\times10^{-9}$ and ${\rm Br}\big( \tau\to\mu\gamma \big)_{\rm Belle\,II}<6.9\times10^{-9}$ are expected from this upgrade~\cite{Belle2lfvestimation}.
\\

The mystery of whether neutrinos were massive prevailed for quite some time, since their introduction, in 1930, until the first measurements of neutrino oscillations, at the Super-Kamiokande and at the Sudbury Neutrino Observatory, reported in 1998 and 2002~\cite{nuoscillationsSKamiokande,nuoscillationsSNO}. Neutrino oscillations is a quantum phenomenon by which the probability of measuring a neutrino with a lepton flavor different from the one that originally characterized it is nonzero, after the neutrino has traveled across some distance from its source.~\cite{Pontecorvo}. Massiveness of neutrinos is a necessary condition for neutrino oscillations to occur, so the observation of this phenomenon has been interpreted as solid evidence supporting nonzero neutrino masses. Clearly, this experimental fact contradicts the Standard Model, where neutrinos are massless, and thus incarnates a manifestation of new physics. While valuable data on quadratic neutrino-mass differences $\Delta m_{jk}^2=m_{n_j}^2-m_{n_k}^2$ have been extracted from several experimental facilities focused on measurements of neutrino oscillations~\cite{KamLANDdmass,SKamiokandedmass1,SKamiokandedmass2,RENOdmass,MINOSdmass,NOvAdmass,IceCubedmass,T2Kdmass,DayaBaydmass}, the absolute neutrino-mass scale cannot be determined by this mean. An upper limit on the sum of light-neutrino masses as stringent as $\sum_jm_{n_j}<0.12\,{\rm eV}$ has been determined, at $95\%\,{\rm C.L.}$, from cosmological observations~\cite{ApacheObservatorynumass,Plancknumass}. Moreover, under the assumption that neutrinos are Majorana fermions, and in view of the lack of measurements of the elusive neutrinoless double beta decay~\cite{nohayndbd,CUPIDMOndbd,CUOREndbd,GERDAndbd,Majoranandbd,EXO200ndbd,KamLANDZenndbd}, estimations of upper bounds on the neutrino effective mass $m_{\beta\beta}=\big| \sum_j\big( U_{\rm PMNS} \big)_{ej}^2m_{n_j} \big|$, lying within the $\sim10^{-2}\,{\rm eV}-10^{-1}\,{\rm eV}$ energy range, have been established through exploration of different isotopes~\cite{CUOREndbd,KamLANDZenndbd,GERDAndbd}. A constraint on the effective electron anti-neutrino mass $m_{\nu_e}^{2({\rm eff})}=\sum_j|\big( U_\nu \big)_{ej}|^2m_{\nu_j}^2$, established by the KATRIN Collaboration~\cite{KATRIN}, has been translated into the upper limit $0.8\,{\rm eV}$ on neutrino mass. While the KATRIN result is not the most stringent constraint, a worth comment on this bound regards its generality, as it is independent on cosmological assumptions and on whether neutrinos are Dirac or Majorana. In what follows, we take the neutrino-mass upper bound by KATRIN as reference. Neutrino masses, in normal hierarchy (NH) or inverted hierarchy (IH), can be related to quadratic mass differences as
\begin{equation}
\textrm{NH}
\left\{
\begin{array}{l}
	m_1=\sqrt{m_3^2-\Delta m_{31}^2},
	\vspace{0.3cm} \\
	m_2=\sqrt{m_3^2-\Delta m_{32}^2},
\end{array}
\right.
\label{NHnumass}
\end{equation}
%%%%%%%%%%
\begin{equation}
\textrm{IH}
\left\{
\begin{array}{l}
	m_1=\sqrt{m_2^2-\Delta m_{21}^2},
	\vspace{0.3cm} \\
	m_3=\sqrt{m_2^2+\Delta m_{32}^2},
\end{array}
\right.
\end{equation}
where either $\sqrt{\Delta m_{31}^2} \leqslant m_3 <0.8\,{\rm eV}$, in the NH scheme, or $\sqrt{-\Delta m_{32}^2} \leqslant m_2<0.8\,{\rm eV}$, if IH is considered. The PDG recommends the values~\cite{PDG}
\begin{equation}
\textrm{NH}
\left\{
\begin{array}{l}
	\Delta m_{21}^2=\big( 7.53\pm0.18 \big)\times10^{-5}\,{\rm eV}^2,
	\vspace{0.3cm} \\
	\Delta m_{32}^2=\big( 2.455\pm0.028 \big)\times10^{-3}\,{\rm eV}^2,
\end{array}
\right.
\end{equation}
%%%%%%%%%%\begin{equation}
\begin{equation}
\textrm{IH}
\left\{
\begin{array}{l}
	\Delta m_{21}^2=\big( 7.53\pm0.18 \big)\times10^{-5}\,{\rm eV}^2,
	\vspace{0.3cm} \\
	\Delta m_{32}^2=\big( -2.529\pm0.029 \big)\times10^{-3}\,{\rm eV}^2,
\end{array}
\right.
\end{equation}
for the light-neutrino quadratic-mass differences $\Delta m_{21}^2$ and $\Delta m_{32}^2$. We have observed that our results do not appreciably change if either of the light-neutrino mass orderings is assumed, so from here on all our estimations are carried out by taking the NH.
\\

We define the dimensionless ratio $x_\sigma\equiv\frac{v_\sigma^2}{m_W^2}$, and then we consider the electromagnetic-moment contributions $\mu^{\beta\alpha}$ and $d^{\beta\alpha}$, shown in Eqs.~(\ref{mubetaalpha}) and (\ref{dbetaalpha}), taking into account the inverse seesaw energy-scale hierarchy $v_\sigma\ll v\ll\Lambda$, previously discussed. Recall that $M\sim\Lambda$ and $\mu_S\sim v_\sigma$. We start by taking the limits of $\mu^{\beta\alpha}$ and $d^{\beta\alpha}$ as $x_\sigma\to0$, that is: 
\begin{eqnarray}
\lim_{x_\sigma\to0}\Big(
\sum_{j=1}^3{\cal B}_{\beta n_j}{\cal B}_{\alpha n_j}^*\mu^{\beta\alpha}_{n_j}+\sum_{j=1}^6{\cal B}_{\beta f_j}{\cal B}_{\alpha f_j}^*\mu^{\beta\alpha}_{f_j}
\Big),
\\
\lim_{x_\sigma\to0}\Big(
\sum_{j=1}^3{\cal B}_{\beta n_j}{\cal B}_{\alpha n_j}^*d^{\beta\alpha}_{n_j}+\sum_{j=1}^6{\cal B}_{\beta f_j}{\cal B}_{\alpha f_j}^*d^{\beta\alpha}_{f_j}
\Big).
\end{eqnarray}
In first place, notice that , in accordance with Eq.~(\ref{ISSformula}), $x_\sigma\to0$ leads to $x_{n_j}\to0$, where $x_{n_j}=\frac{m_{n_j}^2}{m_W^2}$, corresponding to light-neutrino masses, has been defined. Therefore, $\lim_{x_\sigma\to0}\mu^{\beta\alpha}_{n_j}=\lim_{x_{n_j}\to0}\mu^{\beta\alpha}_{n_j}\equiv\mu^{\beta\alpha}_{\rm light}$ and $\lim_{x_\sigma\to0}d^{\beta\alpha}_{n_j}=\lim_{x_{n_j}\to0}d^{\beta\alpha}_{n_j}\equiv d^{\beta\alpha}_{\rm light}$, where $\mu^{\beta\alpha}_{\rm light}$ and $d^{\beta\alpha}_{\rm light}$ are functions which only depend on charged-lepton masses, $m_\alpha$ and $m_\beta$, and the $W$-boson mass, $m_W$. Also take into account that $\mu^{\beta\alpha}_{f_j}$ and $d^{\beta\alpha}_{f_j}$ are $v_\sigma$-scale independent, so $\lim_{x_\sigma\to0}\mu^{\beta\alpha}_{f_j}=\mu^{\beta\alpha}_{f_j}$ and $\lim_{x_\sigma\to0}d^{\beta\alpha}_{f_j}=d^{\beta\alpha}_{f_j}$. Moreover, Eqs.~(\ref{BnISS}) and (\ref{BfISS}) show that the factors ${\cal B}_{\beta n_j}{\cal B}_{\alpha n_j}^*$ and ${\cal B}_{\beta f_j}{\cal B}_{\alpha f_j}^*$ do not depend on the $\mu_S$ matrix, so the limit as $x_\sigma\to0$ is innocuous to them as well. Regarding the first of these factors, note that if $\eta$ is assumed to be small, which according to Eq.~(\ref{nonunitaritymatrix}) is assured as long as the energy-scale $\Lambda$ is large, then $\sum_{j=1}^3{\cal B}_{\beta n_j}{\cal B}_{\alpha n_j}^*\simeq-2\eta_{\beta\alpha}$, for $\alpha\ne\beta$, which holds in our case. With this in mind, the afore-considered limits are given by
\begin{eqnarray}
\lim_{x_\sigma\to0}\mu^{\alpha\beta}=-2\mu^{\beta\alpha}_{\rm light}\,\eta_{\beta\alpha}+\sum_{j=1}^6{\cal B}_{\beta f_j}{\cal B}_{\alpha f_j}^*\,\mu^{\beta\alpha}_{f_j},
\\
\lim_{x_\sigma\to0}d^{\alpha\beta}=-2d^{\beta\alpha}_{\rm light}\,\eta_{\beta\alpha}+\sum_{j=1}^6{\cal B}_{\beta f_j}{\cal B}_{\alpha f_j}^*\,d^{\beta\alpha}_{f_j}.
\end{eqnarray}
Let us analyze these equations in the context of a very large energy-scale $\Lambda$, which translates into very large masses for the heavy neutral leptons. Under such circumstances, we find that the dependence of $\mu^{\beta\alpha}_{f_j}$ and $d^{\beta\alpha}_{f_j}$ on the masses of the heavy neutral leptons $f_j$ is marginal indeed, so $\mu^{\beta\alpha}_{f_j}\simeq\mu^{\beta\alpha}_{\rm heavy}$ and $d^{\beta\alpha}_{f_j}\simeq d^{\beta\alpha}_{\rm heavy}$, which, as the notation indicates, are determined solely by charged-lepton and $W$-boson masses. Then note that $\sum_{j=1}^6{\cal B}_{\beta f_j}{\cal B}_{\alpha f_j}^*=2\eta_{\beta\alpha}$ is fulfilled, so we find the expressions
\begin{equation}
\tilde\mu^{\beta\alpha}=
\hat\mu^{\beta\alpha} \eta_{\beta\alpha},
\label{limitmu}
\end{equation}
%%%%%%%%%%
\begin{equation}
\tilde{d}^{\beta\alpha}=
\hat{d}^{\beta\alpha} \eta_{\beta\alpha},
\label{limitd}
\end{equation}
where $\hat\mu^{\beta\alpha}=2\big(
\mu^{\beta\alpha}_{\rm heavy}-\mu^{\beta\alpha}_{\rm light}
\big)$ and $\hat d^{\beta\alpha}=2\big(
d^{\beta\alpha}_{\rm heavy}-d^{\beta\alpha}_{\rm light}
\big)$. 
%Eqs.~(\ref{limitmu}) and (\ref{limitd}) are both given in terms of the $\alpha\beta$-th entry of $\eta$. 
About Eqs.~(\ref{limitmu})-(\ref{limitd}), note that if, rather than just assuming the energy-scale $\Lambda$ to be large, the formal limit as $\Lambda\to\infty$ is taken, then the factors $\eta_{\beta\alpha}$, in both expressions, must be replaced by $\lim_{x_\Lambda\to\infty}\eta_{\beta\alpha}$, with $x_\Lambda=\frac{\Lambda^2}{m_W^2}$. Following the definition of the matrix $\eta$, given in Eq.~(\ref{nonunitaritymatrix}), this limit is 0, in which case both the $\mu^{\beta\alpha}$ and the $d^{\beta\alpha}$ vanish, which shows the decoupling of the new physics at large energy $\Lambda$. In relation with that, note that ${\cal B}_n$ becomes unitary whereas ${\cal B}_f$ vanishes. Getting back to Eqs.~(\ref{limitmu})-(\ref{limitd}), the explicit expressions of $\hat\mu^{\beta\alpha}$ and $\hat d^{\beta\alpha}$, in terms of masses $m_\alpha$, $m_\beta$, and $m_W$, read
\begin{eqnarray}
&&
\hat{\mu}_{\alpha \beta} = \frac{\hat\chi_{\alpha\beta}}{m_\alpha+m_\beta}\Big( -2m_\alpha^2(m_\alpha+m_\beta)^2
\nonumber \\ && \hspace{0.4cm}
+3\frac{m_\alpha}{m_\beta}(m_\alpha m_\beta-m_W^2) (m_\alpha m_\beta+2 m_W^2)
\nonumber \\ && \hspace{0.4cm}
+3(m_W^2-m_\alpha^2)\Big( m_\beta^2-2\frac{m_W^4}{m_\alpha^3} (2 m_\alpha+m_\beta)
\nonumber \\ && \hspace{0.4cm}
+\frac{m_W^2}{m_\alpha^2} (2 m_\alpha+m_\beta) (2 m_\alpha+3 m_\beta)\Big) \log \bigg(\frac{m_W^2}{m_W^2-m_\alpha^2}\bigg)
\nonumber \\ && \hspace{0.4cm}
+6 m_W^2m_\alpha^2 (-2 m_\alpha^2-3 m_\alpha m_\beta - 2 m_\beta^2+2 m_W^2)C_0
\Big)
\nonumber \\ && \hspace{0.4cm}
+\big( \alpha\longleftrightarrow\beta \big),
\label{muinlimit}
\end{eqnarray}
%%%%%%%%%%
\begin{eqnarray}
&&
\hat{d}_{\alpha \beta}= \frac{\hat\chi_{\alpha\beta}}{m_\alpha-m_\beta} \Big(-2m_\alpha^2(m_\alpha-m_\beta)^2
\nonumber \\ && \hspace{0.4cm}
-3 \frac{m_\alpha}{m_\beta}(m_\alpha m_\beta+ m_W^2)(m_\alpha m_\beta -2 m_W^2)
\nonumber \\ && \hspace{0.4cm}
+3(m_W^2-m_\alpha^2)\Big( m_\beta^2-2\frac{m_W^4}{m_\alpha^3}(2 m_\alpha-m_\beta)
\nonumber \\ && \hspace{0.4cm}
+\frac{m_W^2}{m_\alpha^2}(2m_\alpha-m_\beta)(2m_\alpha-3m_\beta)\Big) \log \bigg(\frac{m_W^2}{m_W^2-m_\alpha^2}\bigg)
\nonumber \\ && \hspace{0.4cm}
+6m_W^2m_\alpha^2(-2 m_\alpha^2+3 m_\alpha m_\beta-2 m_\beta^2+2 m_W^2) C_0
\Big)
\nonumber \\ && \hspace{0.4cm}
-\big( \alpha\longleftrightarrow\beta \big),
\label{dinlimit}
\end{eqnarray}
where we have defined
\begin{equation}
\hat\chi_{\alpha\beta}=\frac{e^3}{3(8\pi)^2s_{\rm W}^2m_W^2 (m_\alpha^2-m_\beta^2)},
\end{equation}
with $s_{\rm W}=\sin\theta_{\rm W}$ the sine of the weak-mixing angle, whereas the short-hand notation
\begin{equation}
C_0=C_0(0,m_\alpha^2,m_\beta^2,m_W^2,m_W^2,0)
\end{equation}
has been used. As displayed by Eqs.~(\ref{muinlimit}) and (\ref{dinlimit}), the factors $\hat\mu_{\alpha\beta}$ and $\hat d_{\alpha\beta}$ are, respectively, symmetric and antisymmetric with respect to their lepton-flavor indices. In the limit considered for Eqs.~(\ref{limitmu}) and (\ref{limitd}), the branching ratio for $l_\alpha\to\,l_\beta\gamma$, previously given in Eq.~(\ref{exactBr}), is expressed as
\begin{equation}
{\rm Br}\big( l_\alpha\to\,l_\beta\,\gamma \big)=\frac{\big( m_\alpha^2-m_\beta^2 \big)^3}{8\pi m_\alpha^3\Gamma_{\rm tot.}}\Big( |\hat\mu_{\beta\alpha}|^2+|\hat d_{\beta\alpha}|^2 \Big)|\eta_{\beta\alpha}|^2.
\label{Brinlimit}
\end{equation}
From this equation, the link among off-diagonal non-unitary effects $\eta_{\beta\alpha}$, with $\beta\ne\alpha$, and the branching ratios for lepton-flavor-changing decays $l_\alpha\to\,l_\beta\,\gamma$ can be appreciated. 
\\

The matrix $\eta$ is Hermitian, so only six of its entries are independent, with a total number of 9 parameters. Neutrino oscillations is a mean to search for non-unitary traces. For instance, Ref.~\cite{FGTT} presents a study in which joint data from short-baseline and long-baseline neutrino-oscillations experiments are used to constrain parameters quantifying non-unitarity. The analyses performed by the authors of that paper lead to upper limits on $|\eta_{\alpha\beta}|$ as stringent as $10^{-2}$. Future experimental facilities, aimed at high-precision studies of neutrino oscillations, are also available, though notice that, in the most optimistic cases, these machines will be able to set upper bounds on non-unitarity parameters as restrictive as $\sim10^{-3}$~\cite{EFMTV,BCFHL,MPSTV,Soumya,ADGM,CEFM}. A dedicated global fit, comprehending several scenarios in which heavy-neutral leptons are added to the Standard-Model particle content, was carried out by the authors of Ref.~\cite{FHL}. An ulterior work~\cite{BFHLMN}, which features the same authors and further collaborators, has improved and updated this analysis, establishing the preferable regions
\begin{eqnarray}
|\eta_{e\mu}|<1.2\times10^{-5},
\label{etaemu}
\\
|\eta_{e\tau}|<8.8\times10^{-4},
\\
|\eta_{\mu\tau}|<1.8\times10^{-4},
\end{eqnarray}
at $95\%\,{\rm C.L.}$, on the non-unitarity parameters which are related to the present investigation. To perform our estimations, the matrix $\hat{V}$, first introduced in Eq.~(\ref{metadiagonalization}), is related to the PMNS matrix as $\hat{V}^*=U_{\rm PMNS}$. The set of parameters defining the PMNS neutrino-mixing matrix, $U_{\rm PMNS}$, is therefore another aspect to take into account for our estimations. The number of such parameters depends on whether the neutrinos are described by Dirac or Majorana fields. In the case of Majorana neutrinos, this matrix is expressed as $U_{\rm PMNS}=U_{\rm D}U_{\rm M}$, where $U_{\rm M}$ is a $3\times3$ diagonal matrix, given in terms of the so-called Majorana phases, $\phi_1$ and $\phi_2$, as $U_{\rm M}={\rm diag}\big( 1,e^{i\phi_1},e^{i\phi_2} \big)$. For the sake of simplicity, from here on we take the values $\phi_1=0$ and $\phi_2=0$ for the Majorana phases. The other matrix factor in $U_{\rm PMNS}$ is written as
\begin{widetext}
\begin{equation}
U_{\rm D}=
\left(
\begin{array}{ccc}
	c_{12}c_{13} & s_{12}s_{13} & s_{13} e^{-i\delta_{\rm D}}
	\vspace{0.2cm}
	\\
	-s_{12}c_{23}-c_{12}s_{23}s_{13}e^{i\delta_{\rm D}} & c_{12}c_{23}-s_{12}s_{23}s_{13}e^{i\delta_{\rm D}} & s_{23}c_{13}
	\vspace{0.2cm}
	\\
	s_{12}s_{23}-c_{12}c_{23}s_{13}e^{i\delta_{\rm D}} & -c_{12}s_{23}-s_{12}c_{23}s_{13}e^{i\delta_{\rm D}} & c_{23}c_{13}
\end{array}
\right),
\end{equation}
\end{widetext}
where mixing angles $\theta_{12}$, $\theta_{23}$, and $\theta_{13}$ reside within trigonometric functions, briefly denoted as $s_{jk}=\sin\theta_{jk}$ and $c_{jk}=\cos\theta_{jk}$, whereas a $CP$-violating complex phase, given the name ``Dirac phase'', is denoted by $\delta_{\rm D}$. Using experimental data by a number of experimental collaborations, the Particle Data Group (PDG) reports the following values for the mixing angles~\cite{PDG}:
\begin{eqnarray}
&&
\sin^2\theta_{12}=0.307\pm0.013,
\label{s12squared}
\\ &&
\sin^2\theta_{23}=0.546\pm0.0021,
\label{s23squared}
\\ &&
\sin^2\theta_{13}=0.0220\pm0.0007.
\label{s13squared}
\end{eqnarray}
The PDG considered Ref.~\cite{SKamiokandedmass1}, by the Super-Kamiokande Collaboration, to set the $s_{12}^2$ value shown in Eq.~(\ref{s12squared}). For $s_{23}^2$, Eq.~(\ref{s23squared}), the PDG took the results by Super-Kamiokande~\cite{SKamiokandedmass2}, Minos+~\cite{MINOSdmass}, T2K~\cite{T2Ks23},  NOvA~\cite{NOvAs23}, and IceCube~\cite{IceCubes23} . The results followed by the PDG to give the value of $s_{13}^2$, which is displayed in Eq.~(\ref{s13squared}), were provided by DayaBay~\cite{1DayaBays13,2DayaBays13}, RENO~\cite{RENOdmass,RENOs13}, and Double Chooz~\cite{DoubleChoozs13}. A value for the Dirac phase has also been set by the PDG, namely,
\begin{equation}
\delta_{\rm D}=1.23\pm0.21\pi\,{\rm rad},
\label{Diracphase}
\end{equation}
which was obtained on the grounds of the results reported by Super-Kamiokande~\cite{SKamiokandedmass2}, NOvA~\cite{NOvAdmass}, and T2K~\cite{T2Kdmass}. 
\\

\begin{figure}[ht]
\center
\includegraphics[width=9cm]{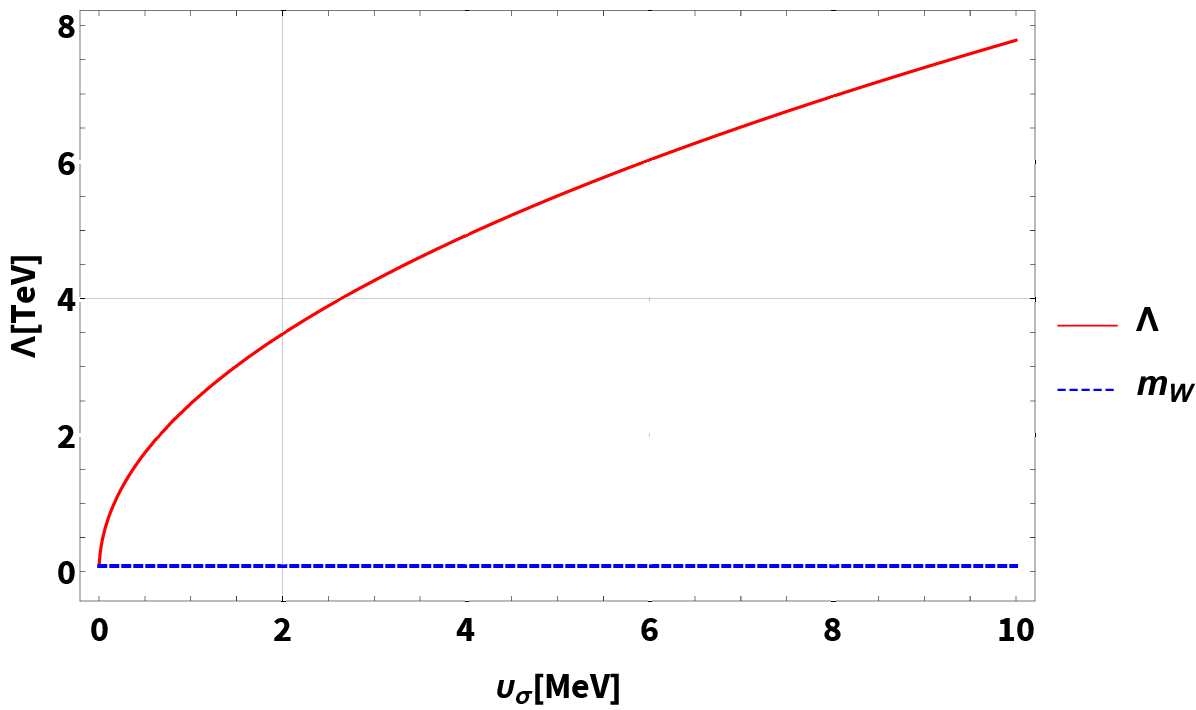}
\caption{\label{Lambdavssigma} Values of the $\Lambda$ scale, for $0\,{\rm MeV}\lesssim v_\sigma\leqslant 10\,{\rm MeV}$, from Eq.~(\ref{scalesrelation}).}
\end{figure}

As we stated above, the $3\times3$ matrix $M$, first introduced in Eq.~(\ref{Yukawamass}), is assumed to be connected with a high-energy scale $\Lambda$, characteristic of some fundamental description of nature, beyond the Standard Model. We assume that such a relation goes as
\begin{equation}
M\simeq\Lambda\,\zeta_M,
\label{MeqsLambda}
\end{equation}
with $\zeta_M$ a diagonal real matrix, in accordance with our previous discussion on the diagonalization of the neutral-lepton mass matrix. Recall that both the $\mu_S$ and the $\mu_R$ matrices, responsible for lepton-flavor violation, have been assumed to be proportional to the scale $v_\sigma$. With this in mind, and for the sake of simplicity, from here on we assume that these matrices are the same, that is, $\mu_S=\mu$ and $\mu_R=\mu$, where the notation $\mu$ is used in a generic manner. We write this matrix as $\mu=\big( \frac{v_\sigma}{\sqrt{2}}\times10^{-2} \big)\zeta_\mu$, with the assumption that $\zeta_\mu$ is a $3\times3$ matrix both real and diagonal. On the other hand, we take~\cite{MSV}	
\begin{equation}
m_{\rm D}=\big(\frac{v}{\sqrt{2}}\times10^{-1}\big)\hat{m}_{\rm D}, 
\label{chidorria}
\end{equation}
with $\hat{m}_{\rm D}$ a $3\times3$ complex matrix. Putting all these ingredients together, we have the light-neutrino mass matrix, Eq.~(\ref{ISSformula}), expressed as
\begin{equation}
M_n=\frac{1}{2\sqrt{2}}\Big( \frac{v^2v_\sigma}{\Lambda^2}\times10^{-4} \Big)\hat{V}^{\rm T}\hat{m}_{\rm D}\,\zeta_M^{-1}\,\zeta_\mu\,\zeta_M^{-1}\,\hat{m}_{\rm D}^{\rm T}\hat{V}.
\label{numassmatrix}
\end{equation}
Furthermore, these assumptions allow us to write the matrix $\eta$, defined in Eq.~(\ref{nonunitaritymatrix}) and which describes non-unitary effects in light-neutrino mixing, as
\begin{equation}
\eta=\Big( \frac{v^2}{4\Lambda^2}\times10^{-2} \Big)\hat{m}_{\rm D}\,\zeta^{-2}_M\,\hat{m}_{\rm D}^\dag.
\end{equation}
A main objective of the present work is the estimation of $\eta$ off-diagonal entries. To this aim, information can be extracted from Eq.~(\ref{numassmatrix}). To begin, let us assume that 
\begin{equation}
\frac{v^2v_\sigma}{\Lambda^2}\times10^{-4}\sim1\,{\rm eV},
\label{scalesrelation}
\end{equation}
which gives, in passing, the value of the high-energy scale $\Lambda$ in terms of $v_\sigma$. In this regard, we provide, in Fig.~\ref{Lambdavssigma},
a graph which shows the comportment of the energy-scale $\Lambda$ with respect to the $v_\sigma$ scale, as dictated by Eq.~(\ref{scalesrelation}). Keep in mind that $\Lambda$ is the energy scale essentially determining the masses of the heavy-neutral leptons, so this graph can be taken as a reference of what to expect about the sizes of such masses for a lepton-number-violating scale ranging within the regime of the few MeVs. Observe that $\Lambda$ achieves values as large as $\sim8\,{\rm TeV}$ for $v_\sigma=10\,{\rm MeV}$. Regarding the matrices $\zeta_{M}$ and $\zeta_\mu$, both of them real and diagonal, we consider values for their in-diagonal entries ranging within $0.5\leqslant\big( \zeta_M \big)_{jj}\leqslant1.5$ and $0.5\leqslant\big( \zeta_\mu \big)_{jj}\leqslant1.5$. We also work under the assumption that $\hat{m}_{\rm D}$ is symmetric. Then, given a set of values for the masses of light neutrinos, the entries of $\hat{m}_{\rm D}$ can be determined from Eq.~(\ref{numassmatrix}), for either the NH or the IH of light-neutrino masses~\cite{PDG}. It turns out that four symmetric matrix-texture solutions for $\hat{m}_{\rm D}$ are determined for each set of fixed light-neutrino masses. 
\\

\begin{figure}[ht]
\center
\includegraphics[width=8cm]{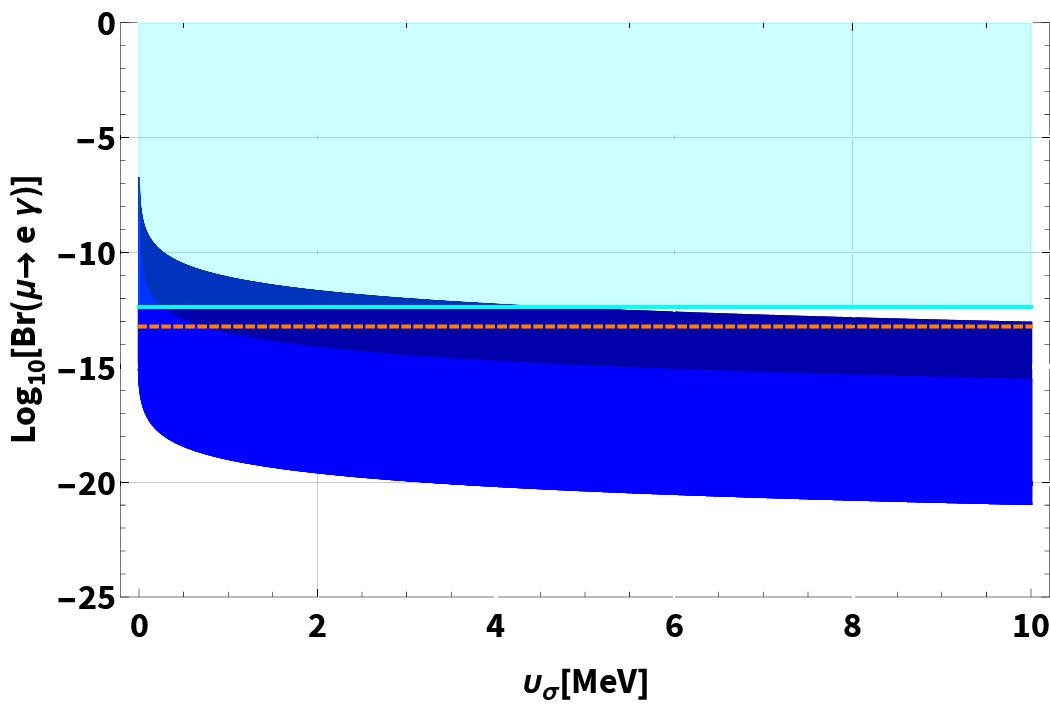}
\vspace{0.2cm}\\ 
\includegraphics[width=8cm]{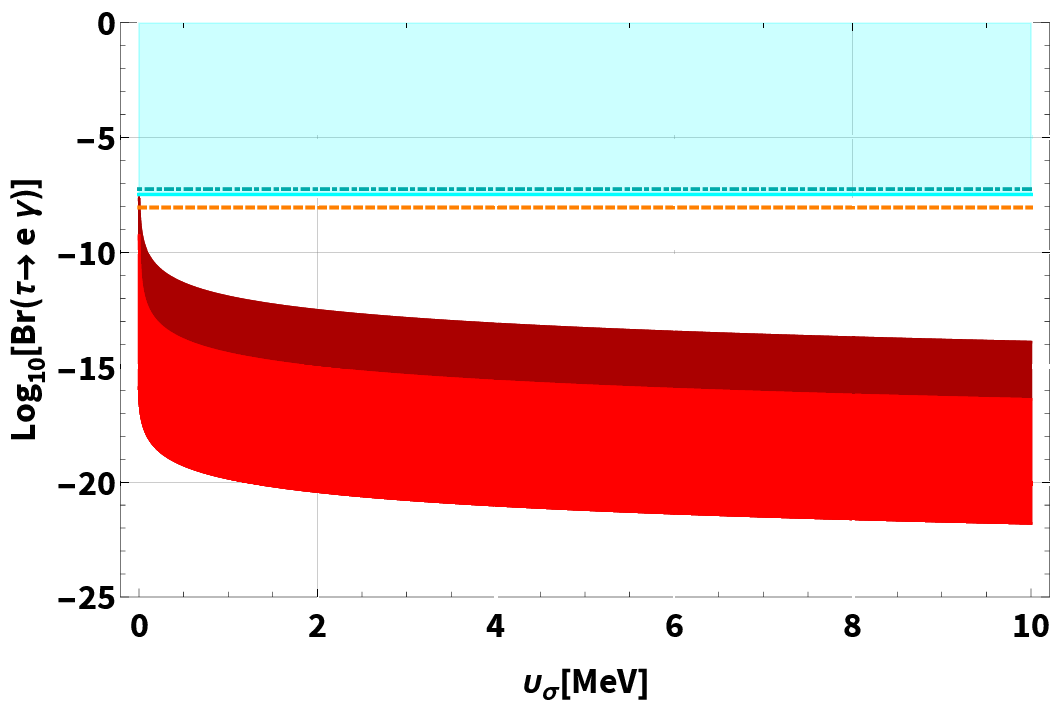}
\vspace{0.2cm}\\
\includegraphics[width=8cm]{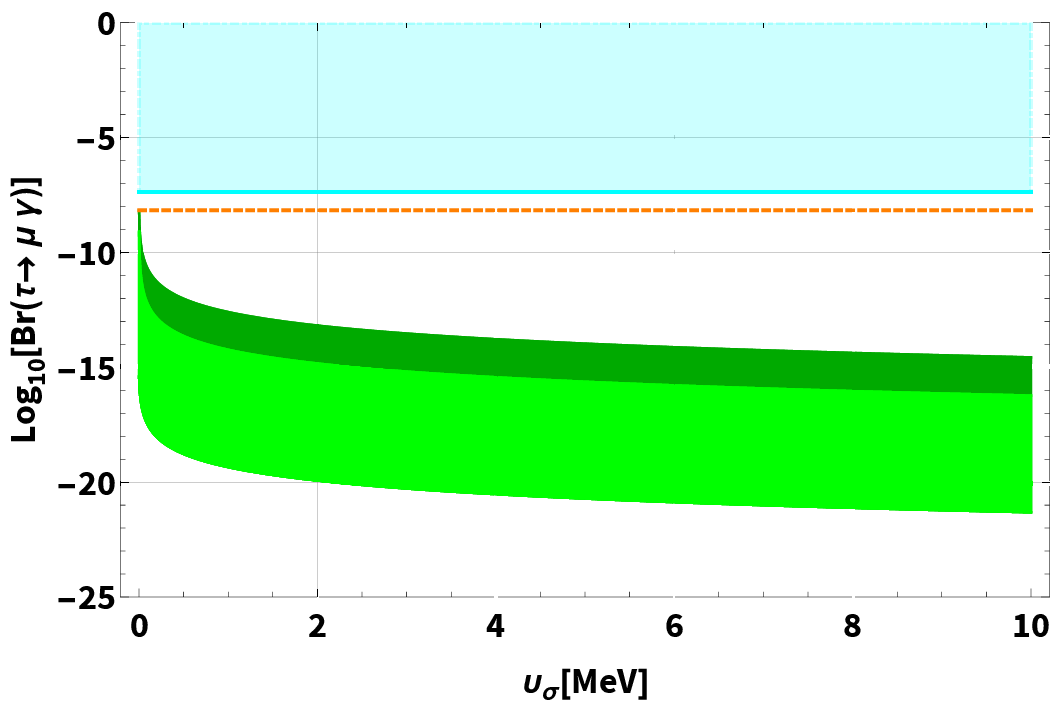}
\caption{\label{BrVSnonu} Contributions from inverse-seesaw massive neutrinos to ${\rm Br}\big( l_\alpha\to l_\beta\gamma \big)$, in base-10 logarithmic scale, as functions of $v_\sigma$. Two regions, one contained within the other, are displayed in each graph: smallest regions for heavy-neutral-lepton quasi-degenerate masses; largest regions for non-degenerate masses of heavy neutral leptons. Horizontal lines represent upper bounds from MEG and MEG II (upper panel), and from Belle, Belle II and BaBar (second and third panels).}
\end{figure}

With the previous discussion in mind, we provide the graphs shown in Fig.~\ref{BrVSnonu},
where ${\rm Br}\big( \mu\to e\gamma \big)$, ${\rm Br}\big( \tau\to e\gamma \big)$, and ${\rm Br}\big( \tau\to\mu\gamma \big)$ are displayed for different values of the scale $v_\sigma$, varied within $1\,{\rm KeV}\leqslant v_\sigma\leqslant10\,{\rm MeV}$. We find it worth emphasizing that such graphs have been made by usage of the exact expression for the branching ratio ${\rm Br}\big( l_\alpha\to l_\beta\,\gamma \big)$, Eq.~(\ref{exactBr}), which, unlike Eq.~(\ref{Brinlimit}), does depend on the masses of both light neutrinos and heavy neutral leptons. Furthermore, while earlier analyses of virtual-neutrino contributions to $\l_\alpha\to l_\beta\,\gamma$, such as those found in Refs.~\cite{GPH,SoRu,IlPi}, use expressions in which approximations regarding both light-neutrino and charged-lepton masses are considered, our numerical estimations have been carried out using exact expressions, given in terms of Passarino-Veltman scalar functions. In order to have a clearer appreciation of the orders of magnitude of the contributions, the values of the branching ratios have been plotted in base-10 logarithmic scale. Note that two shadowed regions have been plotted in each of the three graphs of Fig.~\ref{BrVSnonu}. Any of such shadowed regions represents inverse-seesaw contributions to ${\rm Br}\big( l_\alpha\to l_\beta\gamma\big)$.
%for different values of the parameters $\big( m_3,(\zeta_M)_{jj},(\zeta_{\mu})_{jj} \big)$. 
The difference between the two shadowed regions in each of the graphs is given by assumptions on the heavy-neutral-lepton mass spectrum. More precisely, the smaller shadowed regions correspond to degenerate heavy-neutral-lepton mass spectra, whereas for the elaboration of the larger regions both degenerate and non-degenerate spectra for heavy neutral leptons have been considered. Therefore, in each graph, the smallest of the regions is completely embedded within the other. Let us explain the process of generating the largest shadowed regions of the graphs. As displayed in Eqs.~(\ref{mubetaalpha}) and (\ref{dbetaalpha}), the transition electromagnetic moments $\mu^{\beta\alpha}$ and $d^{\beta\alpha}$, upon which the branching ratio ${\rm Br}\big( l_\alpha\to l_\beta\gamma \big)$ depends, involve, besides the mass-dependent functions $\mu_{n_j}^{\beta\alpha}$, $\mu_{n_j}^{\beta\alpha}$, $d_{n_j}^{\beta\alpha}$, and $d_{n_j}^{\beta\alpha}$, the mixings ${\cal B}_{\alpha n_j}$ and ${\cal B}_{\alpha f_j}$. According to Eqs.~(\ref{BnISS}) and (\ref{BfISS}), such mixings are given by the matrices $m_{\rm D}$, $M^{-1}$, and $\hat{V}^*=U_{\rm PMNS}$. Moreover, from Eq.~(\ref{MeqsLambda}), $M$ is determined by the high-energy scale $\Lambda$ and the $3\times3$ matrix $\zeta_M$, which is both real and diagonal. Therefore, in order to evaluate the mixings ${\cal B}_{\alpha n_j}$ and ${\cal B}_{\alpha f_j}$, something has to be done about the following set of parameters: $U_{\rm PMNS}$, $m_{\rm D}$, $\zeta_M$, $\Lambda$. For $U_{\rm PMNS}$, we have used the values recommended by the Particle Data Group~\cite{PDG}, which are displayed in Eqs.~(\ref{s12squared})-(\ref{Diracphase}). As discussed above, $\Lambda$ is given in terms of the energy scale $v_\sigma$, in accordance with Eq.~(\ref{scalesrelation}), which has been further illustrated in Fig.~\ref{Lambdavssigma}. Note that, as shown in Eq.~(\ref{chidorria}), $m_{\rm D}\propto\hat{m}_{\rm D}$, thus meaning that the $3\times3$ matrix $m_{\rm D}$ is complex and symmetric, but general in any other regard, in which case it introduces a total of 12 parameters. To deal with the matrix $m_{\rm D}$, we used Eq.~(\ref{numassmatrix}), where the light-neutrino mass matrix $M_n$, as well as the real diagonal matrices $\zeta_M$ and $\zeta_\mu$, remain unfixed. From Eq.~(\ref{MeqsLambda}), notice that the in-diagonal entries of $\zeta_M$ determine whether the mass spectrum of the heavy neutral leptons is quasi-degenerate or non-degenerate\footnote{We call the resulting mass spectra ``quasi-degenerate'', and not just ``degenerate'', because the masses of the heavy neutrinos $N_j$ and the neutral leptons $X_j$, though very similar, are not, strictly speaking, the same. However, in practice we take all the masses of the heavy neutral leptons to coincide.}. In general, such in-diagonal entries can differ of each other, which translates into a non-degenerate set of masses. As advised before, general heavy-neutral-lepton mass spectra, not restricted to be degenerate, have been considered for the determination of the largest shadowed regions. We have performed a scan over the parameters constituting $M_n$, $\zeta_M$, and $\zeta_\mu$, giving the largest light-neutrino mass, in the NH, the values\footnote{Recall that once the value of $m_3$ is fixed, the quadratic neutrino-mass differences $\Delta m_{jk}^2$ determine the remaining two light-neutrino masses, as shown in Eq.~(\ref{NHnumass}), for the NH.} $m_3=\sqrt{\Delta m_{31}^2}$ and $m_3=0.8\,{\rm eV}$, and taking the values $(\zeta_M)_{jj}=0.5, 1.5$ and $(\zeta_{\mu})_{jj}=0.5, 1.5$, where $j=1,2,3$. All these values have then been inserted into Eq.~(\ref{numassmatrix}), thus setting, for each set of specific values, a system of equations, which is solved to yield a matrix texture for $m_{\rm D}$. For each configuration defined by specific parameters $\big( m_3,(\zeta_M)_{jj},(\zeta_{\mu})_{jj} \big)$, four solutions for $m_{\rm D}$ are obtained. With the resulting four textures at hand, the branching ratio ${\rm Br}\big( l_\alpha\to l_\beta\gamma \big)$ is then evaluated as a function of $v_\sigma$, from which a set of four curves emerges. We have generated a total of 512 curves per graph. All these curves fall within the largest regions shown in the graphs, which are bounded by two curves. Now consider the smallest regions in the graphs, which, on the other hand, correspond to the case $\zeta_M=\kappa\cdot{\bf 1}_3$, where values for $\kappa$ have been taken to range within $[0.5, 1.5]$. This matrix texture for $\zeta_M$ corresponds to a quasi-degenerate spectrum of masses of the heavy neutral leptons, that is, where $m_{f_j}\approx m_N$, for $j=1,2,3,4,5,6$ and with $m_N$ representing some generic mass. In this context, with less parameters involved, we have been able to generate a total of 1372 curves, all of them falling within the smallest region in each graph. Regarding the upper panel of Fig.~\ref{BrVSnonu}, corresponding to ${\rm Br}\big( \mu\to e\gamma \big)$, a solid horizontal line has been added, which represents the upper bound established by the MEG Collaboration, at $4.2\times10^{-13}$. In the same graph, a dashed horizontal line at $6\times10^{-14}$ is displayed to represent the expected sensitivity for MEG II. An aspect worth pointing out about our estimation is the set of $v_\sigma$-scale values under consideration, which in other works, as it is the case of Refs.~\cite{GPH,SFLYC}, is restricted to lower energy regimes, such as the eVs or the KeVs. Our estimations comprehend KeV-sized values of $v_\sigma$ and then they go all the way to $10\,{\rm MeV}$. As the first graph of Fig.~\ref{BrVSnonu} indicates, it is in the MeV regime where the largest ${\rm Br}\big( \mu\to e\gamma \big)$ contributions cross the MEG and MEG II sensitivities. In this sense, this wider range of $v_\sigma$ values provides valuable information. In the case of the second and third panels of Fig.~\ref{BrVSnonu}, which correspond to the tau decays, the dot-dashed line corresponds to the Belle bound, whereas the solid horizontal line stands for the limit by BaBar. Furthermore, the lower dashed line represents the expected sensitivity for Belle II. Note that the regions for our contributions to the tau decays fall well out of experimental sensitivity. For this reason, from now on we concentrate on the decay process $\mu\to e\gamma$.
\\

\begin{figure}[ht]
\center
\includegraphics[width=8.5cm]{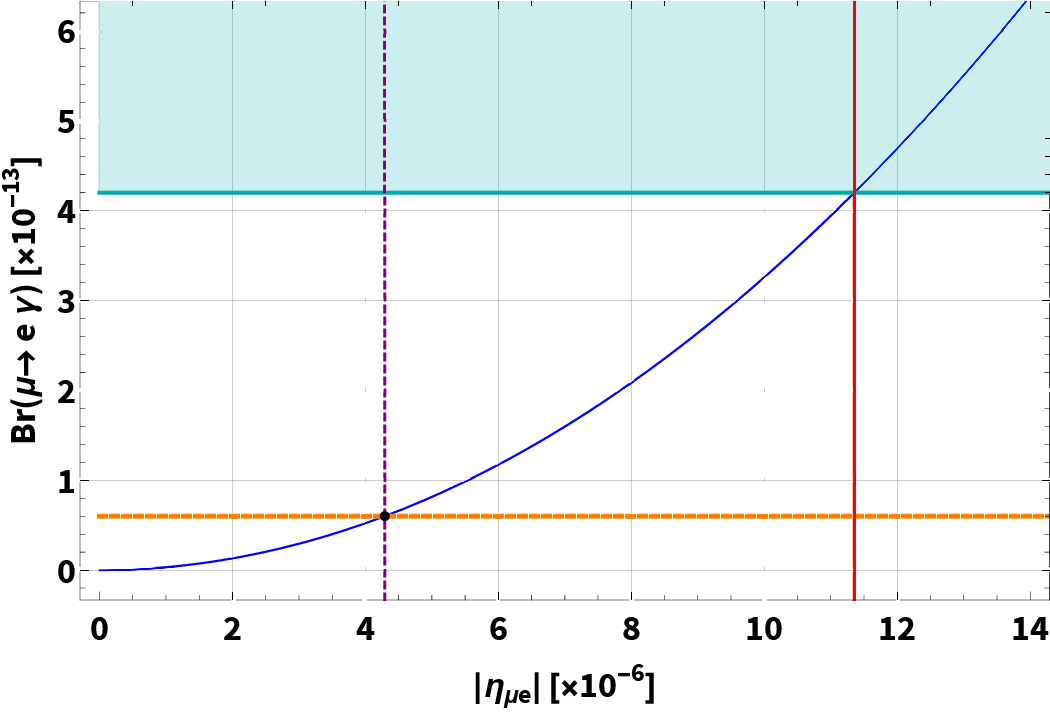}
\caption{\label{nonunitaritygraph} The branching ratio for $\mu\to e\gamma$ as a function of $|\eta_{\mu e}|$, in comparison with the MEG current bound~\cite{MEGlfvbound}, the MEG II expected sensitivity~\cite{MEG2lfvestimation}, and upper limits on $|\eta_{\mu e}|$ in accordance with the preferable region determined in Ref.~\cite{BFHLMN}.}
\end{figure}

Previously, we showed, through Eq.~(\ref{Brinlimit}), how the branching ratio for $l_\alpha\to\,l_\beta\gamma$ is practically determined by the non-unitarity parameters $\eta_{\alpha\beta}$ as far as $v_\sigma\ll v\ll\Lambda$ holds. Such a relation is depicted by the graph of Fig.~\ref{nonunitaritygraph},
where ${\rm Br}\big( \mu\to e\gamma \big)$ is shown as a function of the non-unitary parameter $|\eta_{\mu e}|$, in accordance with Eq.~(\ref{Brinlimit}). Here, we have varied $0<|\eta_{\mu e}|\leqslant14\times10^{-6}$, whereas the comportment of ${\rm Br}\big( \mu\to e\gamma \big)$ within $0$ and $6\times10^{-13}$ is displayed. The solid horizontal line, which bounds the shadowed region from below, represents the MEG upper limit for this branching ratio, whereas the dashed horizontal line corresponds to the expected sensitivity for MEG II. From this graph, we note that the MEG bound on ${\rm Br}\big( \mu\to e\gamma \big)$ yields the constraint $|\eta_{\mu e}|\lesssim1.14\times10^{-5}$, represented by the solid vertical line in the graph. This estimation for $|\eta_{\mu e}|$ is in accordance with the one given by the authors of Ref.~\cite{BFHLMN}, which we show in Eq.~(\ref{etaemu}). Moreover, we observe that a slight improvement of the lower bound on $|\eta_{\mu e}|$ is expected from the MEG II update, which would set the constraint $|\eta_{\mu e}|\lesssim4.29\times10^{-6}$, represented in this graph by the vertical dashed line. As we commented above, constraints on non-unitary effects from experiments of neutrino oscillations are also available, both associated to existing data~\cite{FGTT} and to expected sensitivity of future facilities~\cite{EFMTV,BCFHL,MPSTV,Soumya,ADGM,CEFM}. Also recall that estimated bounds from in-plan devices are expected to be as restrictive as $10^{-3}$.

%%%%%%%%%%%%%%%%%%%%
%%%%%%%%%%%%%%%%%%%%
%%%%%%%%%%%%%%%%%%%%
%%%%%%%%%%%%%%%%%%%%
%%%%%%%%%%%%%%%%%%%%

\section{Summary}
\label{summary}
In this paper we have revisited the contributions from virtual Majorana neutrinos and heavy-neutral leptons to the decay process $l_\alpha\to l_\beta\,\gamma$, at the one-loop level. Such a phenomenon, characterized by lepton-flavor change, is important as its measurement would point towards the presence of new unknown fundamental physics, beyond the Standard Model. To this aim, we have used the inverse seesaw mechanism, for the generation of neutrino masses, as our framework. In this variant of the seesaw, a high-energy scale $\Lambda$, connected to some unspecified fundamental description, is introduced, together with a low-energy scale $v_\sigma$, at which violation of lepton number takes place. These scales and the Higgs-potential vacuum expectation value $v$ fulfill the hierarchy $v_\sigma\ll v\ll\Lambda$. In this context, the inverse-seesaw explanation for tiny light-neutrino masses, currently bounded to be within the sub-eV scale, relies not only on the high-energy scale $\Lambda$, which is the case of the original idea behind the seesaw mechanism, but it is also determined by the scale $v_\sigma$, associated to the violation of lepton number. The version of the inverse seesaw which we considered for the present work introduces a set of six heavy-neutral leptons, which turn out to have very similar masses by pairs. The analytic expression for the $l_\alpha\to l_\beta\,\gamma$ amplitude has been calculated, following the dimensional-regularization approach and the tensor-reduction method. Our results are gauge-invariant, gauge independent, ultraviolet-finite, and well behaved in the sense that the expressions decouple in the limit as $\Lambda\to\infty$. The branching ratios $\mu\to e\gamma$, $\tau\to e\gamma$, and $\tau\to\mu\gamma$ have been estimated and compared with the nowadays best bounds on these decay processes, established by MEG, BaBar and Belle, from which we have concluded that the most promising process among among these is $\mu\to e\gamma$. For the resulting analytic expression of ${\rm Br}(l_\alpha\to l_\beta\,\gamma)$, we have considered the scenario in which $v_\sigma\to0$ and $\Lambda^2\gg m_W^2$, unveiling a direct link among this branching ratio and non-unitary effects in light-neutrino mixing, located in the charged currents featuring the Standard-Model $W$ boson. Using MEG constraint on the branching ratio of the $\mu\to e\,\gamma$ decay, we find results consistent with previous studies on non-unitarity effects and violation of lepton flavor. Moreover, we have observed that projections for MEG II would be able to improve the limit on the non-unitarity parameter $|\eta_{e\mu}|$ by a factor $\sim\frac{1}{3}$. 

%%%%%%%%%%%%%%%%%%%%
%%%%%%%%%%%%%%%%%%%%
%%%%%%%%%%%%%%%%%%%%
%%%%%%%%%%%%%%%%%%%%
%%%%%%%%%%%%%%%%%%%%

\section*{Acknowledgements}
\noindent
We acknowledge financial support from Conahcyt (M\'exico). 

%\appendix*

%\section{}

\end{document}